\documentclass[aps,prmaterials,twocolumn,longbibliography,nobibnotes,showkeys,superscriptaddress]{revtex4-2} 

\usepackage{graphicx}
\usepackage{dcolumn}
\usepackage{bm}
\usepackage{amssymb}
\usepackage{amsmath}

\begin{document}

\title{Native defects, hydrogen impurities, and metal dopants in CeO$_2$}
\author{Khang Hoang}
\email[E-mail: ]{khang.hoang@ndsu.edu}
\affiliation{Center for Computationally Assisted Science and Technology \& Department of Physics, North Dakota State University, Fargo, North Dakota 58108, United States.}
\author{Michelle D. Johannes}
\affiliation{Center for Computational Materials Science, U.S.~Naval Research Laboratory, 4555 Overlook Ave SW, Washington, District of Columbia 20375, United States.}

\date{\today}

\begin{abstract}

Ceria (CeO$_2$) is a material of significant technological importance. A detailed understanding of the material's defect physics and chemistry is key to understanding and optimizing its properties. Here, we report a hybrid density-functional study of native point defects, hydrogen impurities, and metal dopants in CeO$_2$. We find that electron polarons ($\eta_{\rm Ce}^-$) and oxygen vacancies ($V_{\rm O}^{2+}$) are the dominant native defects under conditions ranging from extreme oxidizing to highly reducing. Hydrogen is stable either in the hydroxyl (H$_i^+$) or hydride (H$_{\rm O}^+$) structure but the substitutional H$_{\rm O}^+$ is energetically more favorable than H$_i^+$ only under highly reducing conditions. The interstitial H$_i^+$ is highly mobile in the bulk. Yttrium (Y) is energetically most favorable at the substitutional Ce site. Copper (Cu) and nickel (Ni) can be incorporated at the substitutional site and/or an interstitial site, depending on actual conditions during preparation, and the dopants can exist in different charge and spin states. In light of the results, we discuss electronic and ionic conduction and the effects of metal doping on the formation of electron polarons and oxygen vacancies.

\end{abstract}


\maketitle


\section{Introduction}\label{sec;intro}

Ceria (CeO$_2$), a rare-earth metal oxide, is of interest for numerous important applications. CeO$_2$-based materials have most commonly been used as a catalyst or as a non-inert support for catalysts; they have also been considered for use in fuel cells, hydrocarbon reforming, photocatalysis, water splitting, biomedical applications, among others~\cite{Montini2016CR}. Defect physics and chemistry plays an essential role in the properties and functionalities of materials~\cite{Hoang2018JPCM,Paier2013CR,Schmitt2020CSR}. Oxygen release and uptake in CeO$_2$, for example, arises from the ability to deviate from stoichiometry. Electronic and ionic conduction in the metal oxide is also governed by the formation and migration of respective charge-carrying electronic and ionic defects~\cite{Shoko2011JPCS,Blumenthal1974JES,Tuller1977JPCS,Naik1978JPCS}. In addition to native point defects, which are intrinsic to the material, extrinsic defects such as impurities and dopants are present or intentionally introduced to manipulate the material's properties~\cite{Lee2022NJC,Fierro1987JSSC,Bruce1996ACA,Mao2024IJHE,Zhang2019CC,Ranjith2018SCE,Sartoretti2023CT,DavoQuinonero2020ACSC,Polychronopoulou2021AMI,Wrobel1996JCS,Thurber2007PRB,Derafa2018RSCA,Fuda1984JPCS,Wang1981SSI,Lee2014JPCC}. A detailed understanding of the defect physics and chemistry in CeO$_2$ is key to optimizing its performance in various applications. 

Density-functional theory (DFT) based first-principles studies of defects in CeO$_2$ have been carried out by many authors~\cite{Skorodumova2002PRL,Keating2012JPCC,Zacherle2013PRB,Huang2014JPCC,Plata2013JPCC,Sun2017PRB,Sun2019JPCC,Sun2019AEM,Zhang2023CM,Sohlberg2001JACS,Stimac2025Omega,Wang2010JPCC}, mainly using the DFT$+$$U$ method~\cite{anisimov1991,liechtenstein1995} with the Hubbard $U$ term often applied on the Ce $4f$ states and with few studies using a hybrid DFT/Hartree-Fock method~\cite{heyd:8207,paier:154709}. Although much has been learned from these studies, the fundamental understanding is still far from satisfactory. Certain aspects of defects in CeO$_2$ are under debate; for example, the dominant native defects in the materials under different experimental conditions or the lattice and electronic structure and energetics of specific defects. More importantly, there is still a lack of a more comprehensive and rigorous approach that can provide a deep understanding of the defect physics that in turn can offer physical insights for materials design.      

We herein report a study of native defects, hydrogen impurities, and metal dopants in CeO$_2$, using a hybrid DFT/Hartree-Fock method~\cite{heyd:8207,paier:154709} which treats all orbitals in all elements on equal footing. Supercell models much larger than those used in the previous DFT-based studies are employed to properly take into account local lattice distortion and to reduce artificial defect-defect interaction. In addition to common native defects, hydrogen impurities are selected for this study as CeO$_2$ is often prepared, treated, or used in an H$_2$-rich environment. For metal dopants, we select yttrium (Y), copper (Cu), and nickel (Ni) as examples. These metal dopants are often found in CeO$_2$-based materials~\cite{Zhang2019CC,Ranjith2018SCE,Sartoretti2023CT,DavoQuinonero2020ACSC,Polychronopoulou2021AMI,Wrobel1996JCS,Thurber2007PRB,Derafa2018RSCA,Fuda1984JPCS,Wang1981SSI,Lee2014JPCC} and, as it will be made clear later, they represent rather distinct physics and chemistry. In light of the results for the lattice and electronic structure, energetics, and migration of defects, we discuss possible defect landscapes in CeO$_2$, the electronic and ionic conduction, and the effects of hydrogen impurities and of metal doping. Comparison with previously reported computational studies and with available experiments is made where appropriate. 

\section{Methodology}\label{sec;method}

First-principles calculations are based on the Heyd-Scuseria-Ernzerhof (HSE06) screened hybrid functional \cite{heyd:8207}, the projector augmented wave (PAW) method \cite{PAW1}, and a plane-wave basis set, as implemented in the Vienna {\it Ab Initio} Simulation Package (\textsc{vasp}) \cite{VASP1,VASP2,VASP3}. We use the standard PAW potentials in the \textsc{vasp} database which treat Ce $5s^25p^64f^15d^16s^2$, O $2s^22p^4$, H $1s^1$, Y $4s^24p^64d^15s^2$, Cu $3d^{10}4s^1$, and Ni $3d^84s^2$ explicitly as valence electrons and the rest as core electrons. The Hartree-Fock mixing parameter and the screening length are set to the standard values of 25\% and 10 {\AA}, respectively; the plane-wave basis-set cutoff is set to 500 eV. The calculations for bulk CeO$_2$ (four formula units per unit cell) are carried out using \textbf{k}-point meshes as dense as 8$\times$8$\times$8; \textbf{k}-point meshes for other bulk phases (Ce, Ce$_2$O$_3$, Cu, CuO, Ni, NiO, Y, and Y$_2$O$_3$) and isolated molecules (O$_2$, H$_2$, and H$_2$O) are chosen appropriately. Defects in CeO$_2$ are modeled using cubic 3$\times$3$\times$3 (324-atom) supercells. Integrations over the Brillouin zone in the defect calculations are carried out using the $\Gamma$ point. In all calculations, structural relaxations are performed with the HSE06 functional and the force threshold is chosen to be 0.02 eV/{\AA}; the spin polarization is included and the convergence is assumed when the total energy difference between consecutive cycles is within $10^{-5}$ eV.

The formation energy of a defect or defect complex X in effective charge state $q$ is defined as \cite{walle:3851,Freysoldt2014RMP} 
\begin{align}\label{eq;eform}
E^f({\mathrm{X}}^q)&=&E_{\mathrm{tot}}({\mathrm{X}}^q)-E_{\mathrm{tot}}({\mathrm{bulk}}) -\sum_{i}{n_i\mu_i} \\ %
\nonumber &&+~q(E_{\mathrm{v}}+\mu_{e})+ \Delta^q ,
\end{align}
where $E_{\mathrm{tot}}(\mathrm{X}^{q})$ and $E_{\mathrm{tot}}(\mathrm{bulk})$ are, respectively, the total energies of a supercell containing X and of an equivalent supercell of the perfect host material. $\mu_{i}$ is the atomic chemical potential of species $i$ that have been added to ($n_{i}$$>$0) or removed from ($n_{i}$$<$0) the supercell to form the defect. $\mu_{e}$ is the electron chemical potential, i.e., the Fermi level, referenced to the valence-band maximum (VBM) in the bulk ($E_v$). Finally, $\Delta^q$ is the correction term to align the electrostatic potentials of the perfect bulk and defect supercells and to account for finite-size effects on the total energies of charged defects \cite{Freysoldt}.

We examine defect landscapes in CeO$_2$ using three sets of atomic chemical potentials which correspond to three different experimental conditions. Condition {\bf A} assumes equilibrium with air at $T = 500^\circ$C (within the temperature range in which CeO$_2$ is often prepared), which leads to the oxygen chemical potential $\mu_{\rm O} = -0.87$ eV~\cite{chase}. Condition {\bf B} corresponds to a highly reducing environment in which the host material is assumed to be in equilibrium with the O-deficient phase Ce$_2$O$_3$, which gives $\mu_{\rm O} = -3.09$ eV; equivalently, this condition corresponds approximately to, e.g., the oxygen partial pressure $p_{\rm O_2} = 10^{-10}$ atm and $T = 1200^\circ$C~\cite{chase}. For completeness, we also consider condition {\bf C} which assumes an extreme oxidizing environment where the host is in equilibrium with isolated O$_2$ molecules at 0 K, corresponding to $\mu_{\rm O} = 0$ eV. In each case, the atomic chemical potential of the other species (Ce, H, Y, Cu, and Ni) is determined accordingly using, respectively, CeO$_2$, an isolated H$_2$O molecule, Y$_2$O$_3$, CuO, and NiO as limiting phases, except in the case of {\bf B} where the limiting phase of Ni, Cu, and H is elemental Ni, elemental Cu, and an isolated H$_2$ molecule, respectively. Since {\bf A} and {\bf B} are more experimentally relevant, their results will be explicitly reported; we will mention results obtained under condition {\bf C} only when needed to have a more complete picture.

The migration of a small polaron between two positions $Q_{\rm A}$ and $Q_{\rm B}$ can be described by the transfer of its lattice distortion \cite{Rosso2003}. We estimate the migration barrier ($E_m$) by computing the total energies of a set of supercell configurations linearly interpolated between $Q_{\rm A}$ and $Q_{\rm B}$ and identify the energy maximum. For an oxygen vacancy or hydrogen interstitial, the migration barrier is calculated by using the climbing-image nudged elastic-band (NEB) method \cite{ci-neb}. Unless otherwise noted, these sets of NEB calculations are carried out using the DFT$+$$U$ method \cite{anisimov1991} that is based on the Perdew-Burke-Ernzerhof (PBE) functional \cite{GGA}, with the effective Hubbard parameter $U = 5$ eV applied on the Ce $4f$ states. Castleton et al.~\cite{Castleton2019JPCC} have thoroughly benchmarked various DFT functionals for the study of polaron migration in CeO$_2$ and found that PBE$+$$U$ (with $U = 5$ eV) gives a better description over HSE06. Although the same conclusion may not be made regarding the ionic defects, we also employ PBE$+$$U$ in the NEB calculations for the oxygen vacancy and the hydrogen interstitial to reduce computational costs, thus making the calculations feasible.

Finally, it should be noted that spin-orbit coupling (SOC) is not included as signiﬁcant cancellation is expected between the terms in Eq.~(\ref{eq;eform}). Our tests show that the formation energy of the electron polaron ($\eta^-_{\rm Ce}$) obtained in HSE$+$SOC calculations is different from that obtained in HSE calculations by only 40 meV, well within the typical error bar of the formation energy calculation (about 0.1 eV). This is also consistent with computational tests done for Eu-related defects in GaN~\cite{Hoang2021PRM}.

\section{Results and discussion}\label{sec;results}

\subsection{Bulk properties}\label{subsec;bulk}

\begin{figure}
\vspace{0.2cm}
\centering
\includegraphics[width=\linewidth]{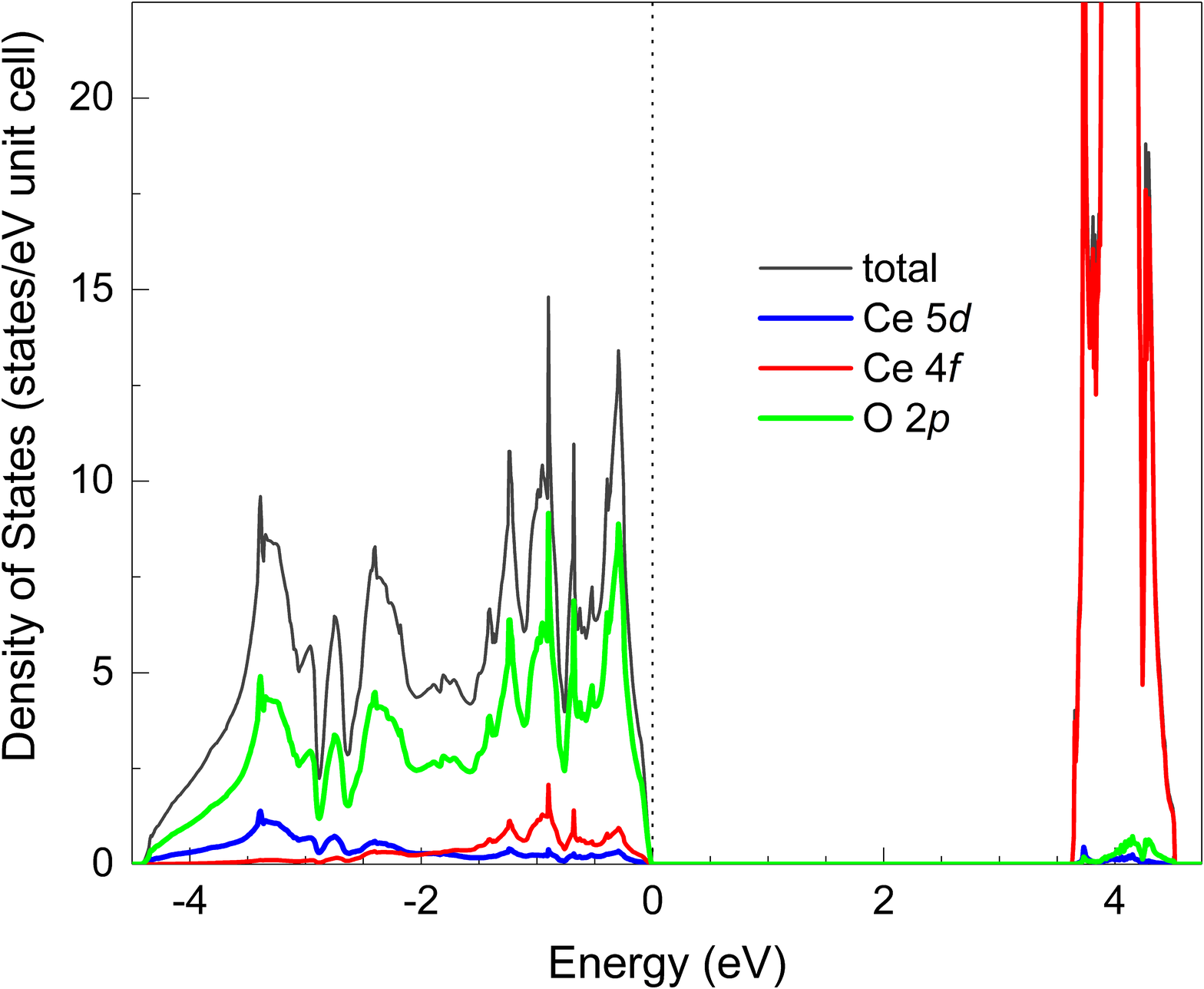}
\vspace{-0.5cm}
\caption{Total and projected electronic densities of states of CeO$_2$. The zero of energy is set to the highest occupied states. }
\label{fig;dos}
\end{figure}

CeO$_2$ crystallizes in a face-centered cubic, fluorite-type structure (space group $Fm\bar{3}m$). The lattice constant within the HSE06 functional is $5.387$ {\AA} (at 0 K), which is comparable to the experimental value ($5.401$ {\AA}, measured at 100 K) \cite{Gupta1970JACeS}. The Ce--O bond length is 2.33 {\AA}. The electronic contribution to the static dielectric constant of CeO$_2$ is 4.56, based on the real part of the dielectric function $\epsilon_{1}(\omega)$ for $\omega\rightarrow0$. The ionic contribution is 21.89, obtained from the macroscopic ion-clamped static dielectric tensor~\cite{Gajdos2006PRB}, as calculated using the \textsc{vasp} code. The total dielectric constant is thus 26.45.

Figure \ref{fig;dos} shows the electronic structure of CeO$_2$. We find that the VBM is predominantly the O $2p$ states, whereas the conduction-band minimum (CBM) is predominantly the Ce $4f$ states. The calculated band gap is 3.62 eV (indirect). For comparison, the reported experimental band gap is in the range 2.9--4.4 eV \cite{Zhang2023CM}. As it will be made clear later, the nature of the electronic structure at the VBM and CBM has great impact on defect formation. Our calculated bulk properties (e.g., band gap and dielectric constant at 0 K) are in agreement with the previous HSE06 bulk calculations~\cite{Lany2024JACS}.

\subsection{Native defects}\label{subsec;defect}  

\begin{figure}
\centering
\includegraphics[width=\linewidth]{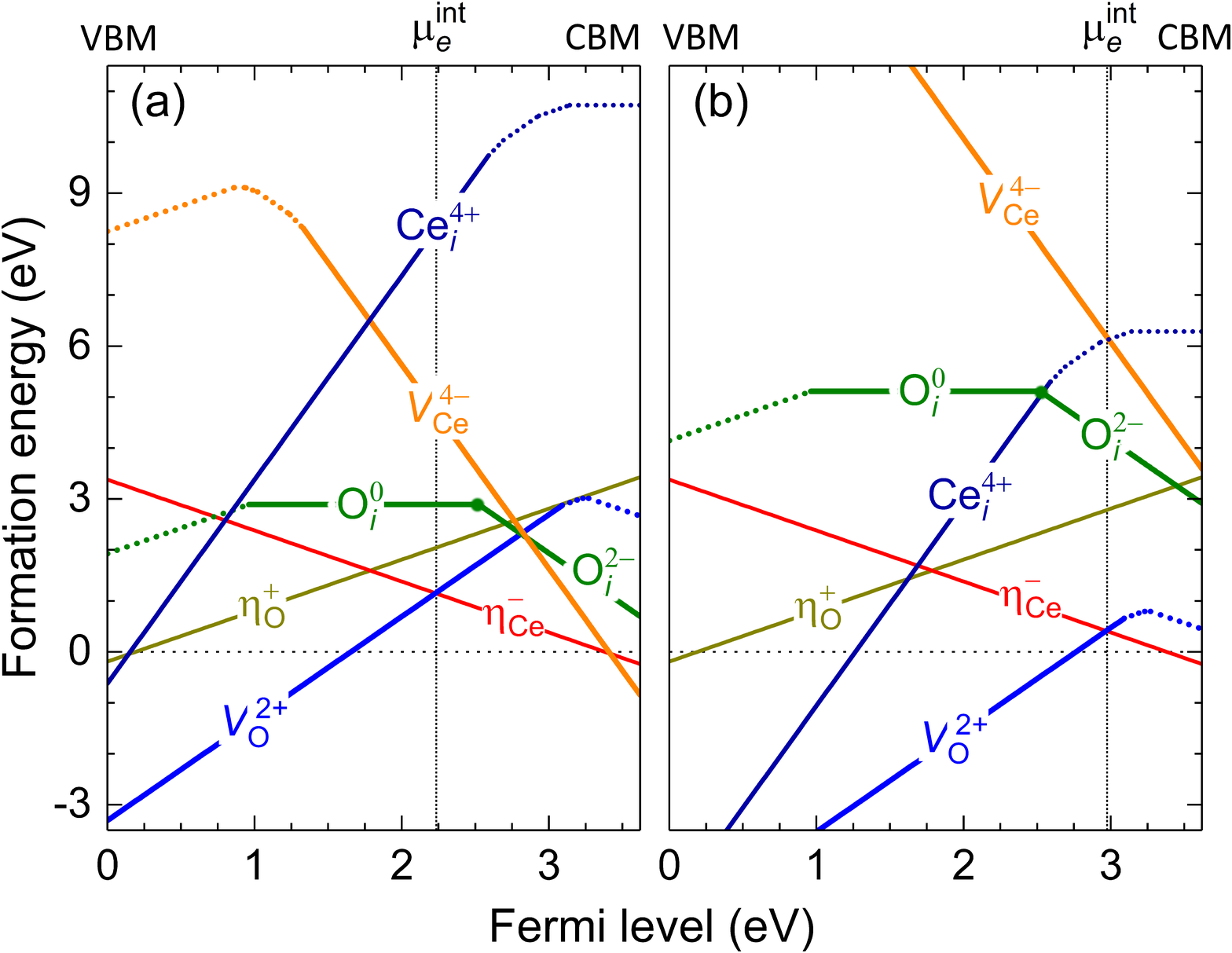}
\vspace{-0.5cm}
\caption{Formation energies of native defects in CeO$_2$ obtained under (a) condition {\bf A} and (b) condition {\bf B}, plotted as a function of the Fermi level from the VBM (at 0 eV) to the CBM (at 3.62 eV). The slope indicates the charge state ($q$): positively (negatively) charged defects have positive (negative) slopes. A solid energy segment represents a stable single point defect configuration; dotted energy segments are complexes of the single defect and one or more electron or hole polarons. $\mu_e^{\rm int}$, marked by the vertical dotted line, is the position of the Fermi level determined by the native defects.}
\label{fig;fe;natives}
\end{figure}  

\begin{figure}
\centering
\includegraphics[width=\linewidth]{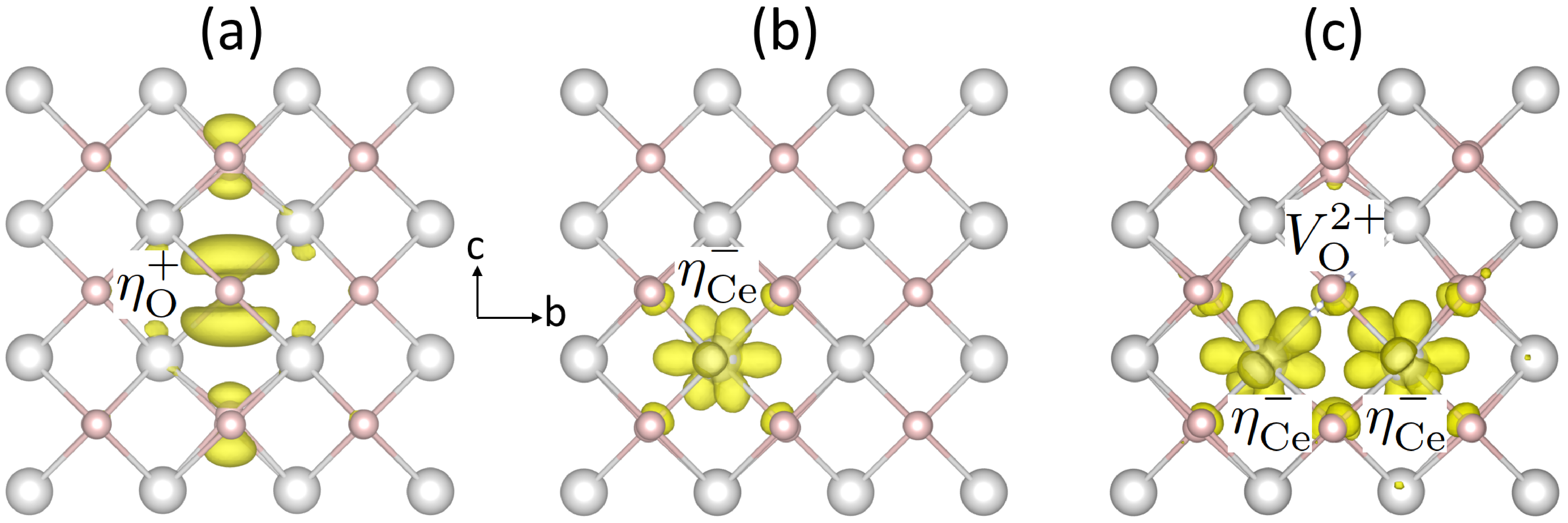}
\vspace{-0.5cm}
\caption{Structures of representative electronic and ionic native defects in CeO$_2$: (a) the hole polaron $\eta_{\rm O}^+$, (b) the electron polaron $\eta_{\rm Ce}^-$, and (c) the neutral oxygen vacancy  (i.e., a complex of $V_{\rm O}^{2+}$ and two $\eta_{\rm Ce}^-$). The isovalue for the charge-density isosurface (yellow) is set to 0.02 e/{\AA}$^{3}$. The large (gray) spheres are Ce and small (red) spheres are O. The vacancy is behind the front oxygen atom at the center. All the atomic structures are visualized using \textsc{vesta}~\cite{VESTA}.}
\label{fig;struct;natives}
\end{figure}    

Figure~\ref{fig;fe;natives} shows the formation energies of relevant native defects in CeO$_2$. Electronic defects include a hole polaron localized mostly at an O site [$\eta^+_{\rm O}$ with spin $S=1/2$, which can be regarded as O$^{-}$ at the O$^{2-}$ site; see Fig.~\ref{fig;struct;natives}(a)] and an electron polaron localized at a Ce site [$\eta^-_{\rm Ce}$, i.e., Ce$^{3+}$ $4f^1$, $S=1/2$, at the Ce$^{4+}$ site; see Fig.~\ref{fig;struct;natives}(b)]. $\eta^+_{\rm O}$ is formed upon removing an electron from the CeO$_2$ supercell; naturally, it is the electron from the highest occupied state (i.e., the VBM, which is predominantly the O $2p$ states) that is removed. $\eta^-_{\rm Ce}$ is, on the other hand, formed upon adding an electron to the supercell, naturally to the lowest unoccupied state (i.e., the CBM, which is predominantly the Ce $4f$ states). These polarons are stable even in the absence of any other defect. The interplay between polaron formation and electronic structure in metal oxides has been well discussed in, e.g., Ref.~\citenum{Hoang2018JPCM}. The local lattice environment is distorted with longer Ce--O bond lengths as the negative (positive) charge gets reduced; 2.42 {\AA} in the case of $\eta^+_{\rm O}$ and 2.41 {\AA} in the case of $\eta^-_{\rm Ce}$, compared to 2.33 {\AA} in the perfect bulk material. Since the hole (electron) is highly localized on the lattice, it can be regarded as small polaron.

Ionic defects include the following structurally, electronically, and energetically stable {\it single} defects: $V_{\rm O}^{2+}$ (i.e., the removal of an O$^{2-}$ ion from the supercell), O$_i^0$ (the addition of an oxygen that leads to the formation of an O--O dimer with an existing oxygen; the O--O distance is 1.42 {\AA}), O$_i^{2-}$ (the added O$^{2-}$ ion is octahedrally coordinated by Ce; the O--Ce distance is 2.55--2.62 {\AA}), $V_{\rm Ce}^{4-}$ (the removal of a Ce$^{4+}$ ion), and Ce$_i^{4+}$ (the addition of a Ce$^{4+}$ ion). In the $V_{\rm O}^{2+}$ configuration, the nearest O neighbors of the vacancy move inward by 0.23 {\AA}, and the nearest Ce neighbors move outward by 0.17 {\AA}. In $V_{\rm Ce}^{4-}$, the nearest O neighbors of the vacancy move outward by 0.23 {\AA}, and the nearest Ce neighbors move inward by 0.07 {\AA}. In Ce$_i^{4+}$, the added Ce$^{4+}$ ion is octahedrally coordinated by Ce with the distance being 3.01--3.02 {\AA}. 

Note that O$_i^0$ at the octahedral site is found to be 1.30 eV higher in energy compared to the O--O dimer configuration. On the other hand, O$_i^{2-}$ in the dimer configuration is 2.45 eV higher compared to being at the octahedral site. Our results for O$_i^{2-}$ appear to be consistent with experiments where the octahedral interstitial site was assumed in structure refinements~\cite{Mamontov2000JPCS,Luo2021CM}. The octahedrally coordinated O$_i^{2-}$ is expected to occur together with $V_{\rm O}^{2+}$ via an anion-Frenkel pair mechanism, and can recombine under high-temperature treatment \cite{Mamontov2000JPCS}.   

Other (nominal) charge states of these vacancies and interstitials are, in fact, {\it defect complexes} consisting of the single defects and the electron or hole polaron(s). For example, the $+$, $0$, or $-$ state of $V_{\rm O}$ is a complex of $V_{\rm O}^{2+}$ and one, two, or three $\eta^-_{\rm Ce}$ with the binding energy of 0.29 eV, 0.44 eV, or 0.55 eV with respect to its constituents, respectively. Figure~\ref{fig;struct;natives}(c) shows the structure of the (nominal) neutral oxygen vacancy with two electron polarons as nearest neighbors. Similarly, the $3-$, $2-$, $-$, $0$, or $+$ state of $V_{\rm Ce}$ is a complex of $V_{\rm Ce}^{4-}$ and one, two, three, four, or five $\eta^+_{\rm O}$ with the binding energy of 1.14 eV, 2.19 eV, 3.01 eV, 3.76 eV, or 4.43 eV, respectively. 
 
Energetically, $\eta^-_{\rm Ce}$ and $V_{\rm O}^{2+}$ have the lowest formation energies and thus are the dominant defects in undoped CeO$_2$, under conditions from extreme oxidizing ($\mu_{\rm O} = 0$ eV) to highly reducing ($\mu_{\rm O} = -3.09$ eV). Note that although $V_{\rm Ce}^{4-}$ has the lowest formation energy in a range of Fermi level values below the CBM, this range is physically inaccessible under experimentally relevant conditions as the formation energy there becomes negative. In the absence of electrically active impurities that can shift the Fermi-level position or when such impurities occur in much lower concentrations than charged native defects, the Fermi level is at $\mu_{e}^{\rm int}$, determined by the native defects (As $\mu_{e}^{\rm int}$ is at least 0.6 eV from the band edges, contributions from free holes and electrons are negligible). $\mu_{e}^{\rm int}$ is the Fermi-level position at which the material maintains its charge neutrality~\cite{Hoang2018JPCM}. In CeO$_2$, $\mu_e^{\rm int}$ is approximately where $\eta^-_{\rm Ce}$ and $V_{\rm O}^{2+}$ have equal formation energies; $\mu_{e}^{\rm int} = 1.95$ eV above the VBM for $\mu_{\rm O} = 0$ eV, 2.24 eV for $\mu_{\rm O} = -0.87$ eV [Fig.~\ref{fig;fe;natives}(a)], and 2.98 eV for $\mu_{\rm O} = -3.09$ eV [Fig.~\ref{fig;fe;natives}(b)]. The Fermi level is thus higher in a more reducing environment, which is consistent with experiments~\cite{Wardenga2016ASS}. The Fermi level was reported to be about 2.7 eV~\cite{Crovetto2016APL} or $\sim$3 eV~\cite{Wardenga2016ASS} above the VBM in CeO$_2$ thin films. Note that our results are different from those of Zacherle et al.~\cite{Zacherle2013PRB}; the discrepancy can be ascribed to the DFT$+$$U$ method used in the previous work in which the $U$ term was applied on the Ce $4f$ orbitals only and all other orbitals were left uncorrected. Our findings are also in contrast to those of Zhang et al.~\cite{Zhang2023CM} (who used a hybrid quantum mechanical/molecular mechanical method) where the anion-Frenkel pair (i.e., $V_{\rm O}^{2+}$ and O$_i^{2-}$) was found to be dominant and determine the Fermi-level position. As it will be made clear later, knowing $\mu_{e}^{\rm int}$ is key to analyzing the interaction between CeO$_2$ and hydrogen impurities or metal dopants.

The formation energy of $\eta^-_{\rm Ce}$ (i.e., Ce$^{3+}$) and $V_{\rm O}^{2+}$ at $\mu_{e}^{\rm int}$ is low, especially under reducing conditions, e.g., 1.14 eV for $\mu_{\rm O} = -0.87$ eV [an oxidizing environment, see Fig.~\ref{fig;fe;natives}(a)] or 0.41 eV for $\mu_{\rm O} = -3.09$ eV [highly reducing, Fig.~\ref{fig;fe;natives}(b)]. With such a low formation energy, positively charged oxygen vacancies and Ce$^{3+}$ can occur with a high concentration. These two defects can be formed simultaneously, e.g., via oxygen loss during materials preparation or under heat treatment in an reducing environment, or during the oxygen release process in electrochemical applications. This is consistent with the fact that CeO$_2$ samples are often O-deficient and contain Ce$^{3+}$ \cite{Tuller1979JES,Nelson2014JACeS}. Using our defect notation, the oxygen release and uptake reaction can be written as
\begin{equation}\label{eq;oxygen}
{\rm O}_{\rm O}^0 + 2{\rm Ce}_{\rm Ce}^0 \rightleftharpoons \frac{1}{2}{\rm O}_2 + V_{\rm O}^{2+} + 2\eta_{\rm Ce}^-.
\end{equation}
The oxygen storage capacity is thus necessarily related to the ability to form oxygen vacancies in the material. 

Notably, among the native defects, $\eta^-_{\rm Ce}$ and $V_{\rm O}^{2+}$ can serve as charge-carrying defects in electronic and ionic conduction, respectively. The migration barrier ($E_m$) of $\eta^-_{\rm Ce}$ is calculated to be 0.19 eV within DFT$+$$U$. The energy is only 43 meV in HSE06 calculations; however, it is known that the hybrid functional can underestimate the polaron migration barrier in metal oxides \cite{Castleton2019JPCC}. In both sets of calculations, the saddle-point configuration has the extra electron almost equally distributed over the two neighboring Ce atoms, which is different from DFT$+$$U$ results of Sun et al.~\cite{Sun2017PRB}, probably due to the smaller (96-atom) supercell used in their calculations. The energy barrier for the migration of $V_{\rm O}^{2+}$ is 0.72 eV. 

Experimentally, electrical conductivity data reported by Blumenthal and Hofmaier~\cite{Blumenthal1974JES} for CeO$_{2-x}$ ($x=0.00424$) clearly shows two regions in the log $\sigma$ {vs.}~$1/T$ plot: a high-$T$ (low-$T$) region with an activation energy ($E_a$) of 0.22 eV (0.59 eV). In the high-$T$ region, $E_a = E_m$, showing that our calculated migration barrier (0.19 eV) is in good agreement with experiment (0.22 eV). The higher activation energy (0.59 eV) measured in the low-$T$ region can be ascribed to defect association; i.e., $E_a = E_m + E_b$, where $E_b = 0.37$ eV is the binding energy. This value is comparable to the binding energy (0.29 eV) of $\eta^-_{\rm Ce}$ and $V_{\rm O}^{2+}$ we mention earlier. Naik and Tien also reported an activation energy of 0.19 eV in the high-$T$ region for CeO$_{2-x}$, independent of $x$ up to $x=0.03$~\cite{Naik1978JPCS}. Note that the formation energy ($E^f$) does not enter the formula for $E_a$ as the charge-carrying defects (athermal $\eta^-_{\rm Ce}$) already pre-exist in the material during the conductivity measurement \cite{Hoang2018JPCM}. Finally, the calculated migration barrier for $V_{\rm O}^{2+}$ is comparable to the experimental activation energy values ($0.64$--$0.82$) for ionic conduction reported for various rare-earth doped CeO$_2$ materials \cite{Wang1981SSI,Huang1998JACeS,Steele2000SSI,Lai2005JACeS}. 

\subsection{Hydrogen impurities}

\begin{figure}
\centering
\includegraphics[width=\linewidth]{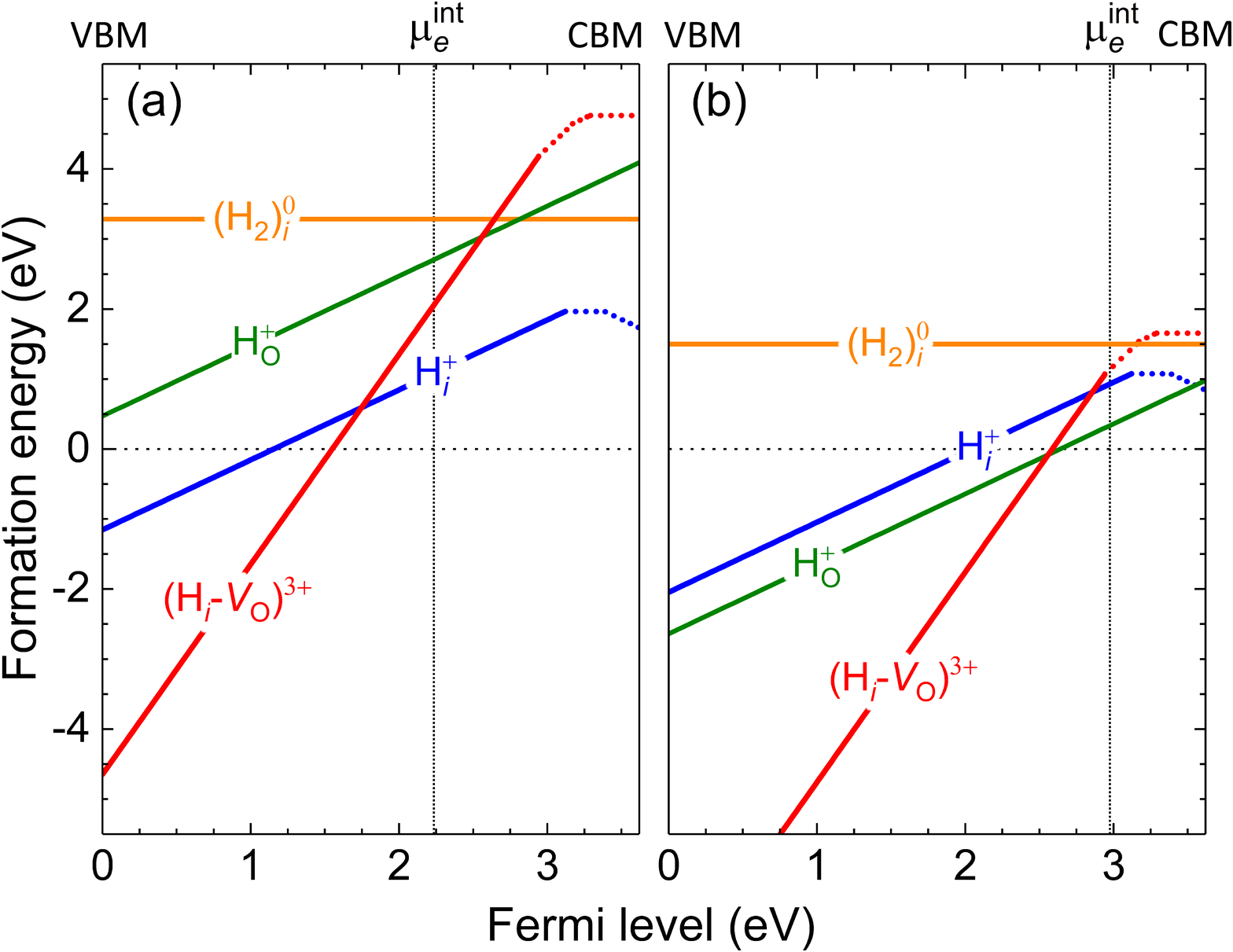}
\vspace{-0.5cm}
\caption{Formation energies of hydrogen impurities in CeO$_2$ obtained under (a) condition {\bf A} and (b) condition {\bf B}. A solid energy segment represents a stable ionic defect configuration; dotted energy segments are complexes of the ionic defect and one or more electron or hole polarons. $\mu_e^{\rm int}$ marks the Fermi-level position determined by the native defects.}
\label{fig;fe;h}
\end{figure}

\begin{figure}
\centering
\includegraphics[width=\linewidth]{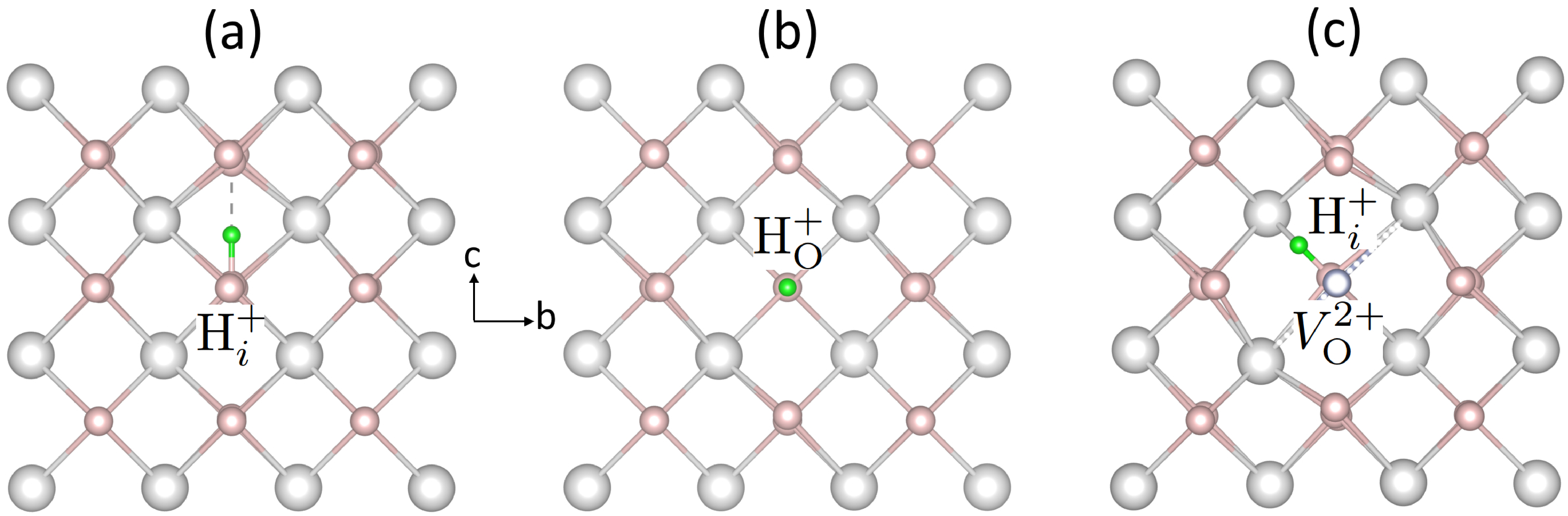}
\vspace{-0.5cm}
\caption{Structures of hydrogen impurities in CeO$_2$: (a) H$_i^+$, (b) H$_{\rm O}^+$, and (c) (H$_i$-$V_{\rm O}$)$^{3+}$ (i.e., a complex of H$_i^+$ and $V_{\rm O}^{2+}$). The small green sphere is H; the circle represents the vacancy.}
\label{fig;struct;h}
\end{figure}   

Figure~\ref{fig;fe;h} shows the formation energies of various hydrogen impurities in CeO$_2$. The hydrogen interstitial is stable as H$_i^+$ with the added proton (H$^+$) forming a {\it hydroxyl} structure with an oxygen (the O--H distance is 1.01 {\AA}) and staying in the line connecting two neighboring O atoms; see Fig.~\ref{fig;struct;h}(a). This H$_i^+$ configuration is 37 meV lower in energy than another one in which the added H$^+$ ion forms a Ce--O--H line (where the O--H distance is only slightly shorter, 0.97 {\AA}). The $0$ (or $-$) state of H$_i$ is not stable as a single defect but a complex of H$_i^+$ and one (two) $\eta_{\rm Ce}^-$ with a binding energy of 0.22 eV (0.25 eV). The hydrogen molecule interstitial is stable as (H$_2$)$_i^0$ with the neutral H$_2$ molecule residing at the center of the Ce octahedron. The substitutional hydrogen at the O lattice site is stable as H$_{\rm O}^+$ which is an H$^-$ ion at the void left by the removal of an O$^{2-}$ ion; see Fig.~\ref{fig;struct;h}(b). Although the H$^-$ ion can be seen as standing alone, its distance to the nearest Ce neighbors is 2.40 {\AA}, comparable to the Ce--H bond lengths in CeH$_3$. H$_{\rm O}^+$ can thus be regarded as having a {\it hydride} structure. Such a substitutional defect can move off-center and form a H$_i$-$V_{\rm O}$ complex. This complex is found to be stable as (H$_i$-$V_{\rm O}$)$^{3+}$, a complex of H$_i^+$ (hydroxyl) and $V_{\rm O}^{2+}$ with a binding energy of 0.18 eV; see Fig.~\ref{fig;struct;h}(c). Other (nominal) charge states of H$_i$-$V_{\rm O}$ are just complexes of (H$_i$-$V_{\rm O}$)$^{3+}$ and one or more $\eta_{\rm Ce}^-$. 

Among these defects, H$_i^+$ is energetically more favorable at $\mu_e^{\rm int}$ for $0$ eV $\leq \mu_{\rm O} \lessapprox -2.50$ eV [which includes condition {\bf A}--see Fig.~\ref{fig;fe;h}(a)--and condition {\bf C}]. Interestingly, H$_{\rm O}^+$ is more favorable for $-2.50$ eV $\lessapprox \mu_{\rm O} \leq -3.09$ eV [which includes condition {\bf B}, see Fig.~\ref{fig;fe;h}(b)]. Under less reducing conditions (which also correspond to lower $\mu_e^{\rm int}$ values), H$_{\rm O}^+$ (hydride) decomposes into H$_i^+$ (hydroxyl) and an oxygen vacancy. Our results thus suggest that one can control the dominant hydrogen species in the bulk by tuning experimental conditions. We also find that H$_i^+$ is highly mobile in the bulk; its migration barrier is only 60 meV within DFT$+$$U$, in agreement with the values recently reported by Stimac and Goldman~\cite{Stimac2025Omega} for (nominally neutral) hydrogen interstitials. The migration of H$^+$ ions from the surface into the bulk may be affected by competing processes, however. The out-diffusion from the bulk to the surface, for example, can counteract the in-diffusion. Besides, hydrogen can react with oxygen on the surface to form H$_2$O or some OH species and get cleaned out. The high activation energy ($<1.69$ eV) for hydrogen diffusion experimentally observed in CeO$_2$ polycrystalline thin films~\cite{Mao2024IJHE} seems to suggest that the in-diffusion may be impeded by certain processes at the surface/subsurface layers or grain boundaries. This issue, especially regarding hydrogen diffusion in CeO$_2$ single crystals, needs further experimental investigations.   

Other experimental reports indicated that hydrogen can be incorporated into the bulk~\cite{Fierro1987JSSC,Bruce1996ACA}, although the solubility appears to be low~\cite{Sakai1999SSI}, and CeO$_2$ prepared under H$_2$ flow has a significantly increased Ce$^{3+}$ concentration~\cite{Wu2017JACS,Lee2022NJC}. The formation of Ce$^{3+}$ is likely associated with the formation of oxygen vacancies. However, as a positively charged defect (donor-like dopant) and {\it if} present in the bulk with a significant concentration, H$_i^+$ can also shift the Fermi level toward the CBM~\cite{Hoang2018JPCM}, thus lowering the the formation energy of $\eta_{\rm Ce}^-$ and therefore increasing its concentration. Finally, Wu et al.~\cite{Wu2017JACS} reported evidence for the presence of bulk Ce--H species upon H$_2$ dissociation over CeO$_2$. Such hydride species could be related to the H$_{\rm O}^+$ defect we discussed above.

\subsection{Metal dopants}

\begin{figure}
\centering
\includegraphics[width=\linewidth]{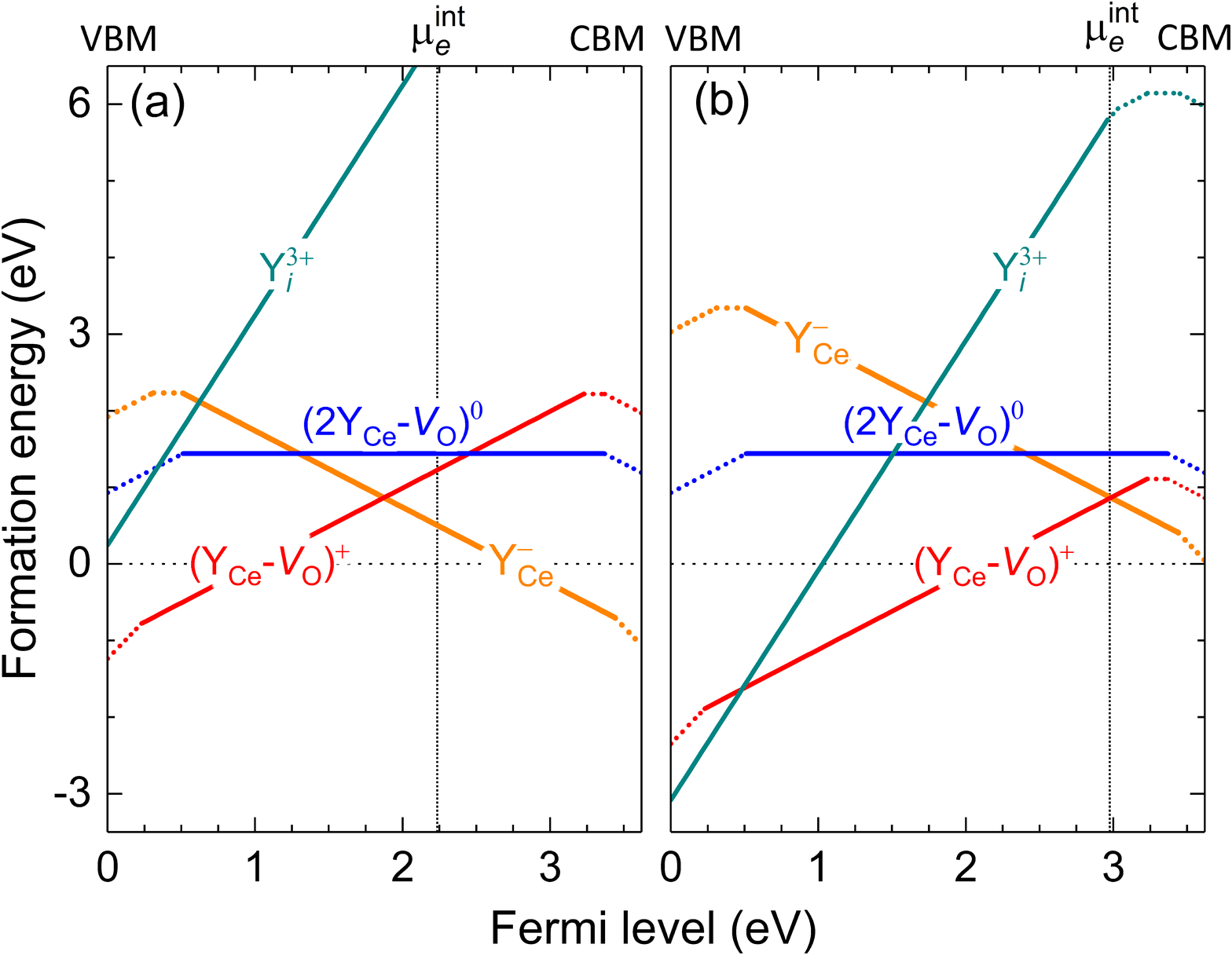}
\vspace{-0.5cm}
\caption{Formation energies of Y-related defects in CeO$_2$ obtained under (a) condition {\bf A} and (b) condition {\bf B}.}
\label{fig;fe;y}
\end{figure}

\begin{figure}
\centering
\includegraphics[width=\linewidth]{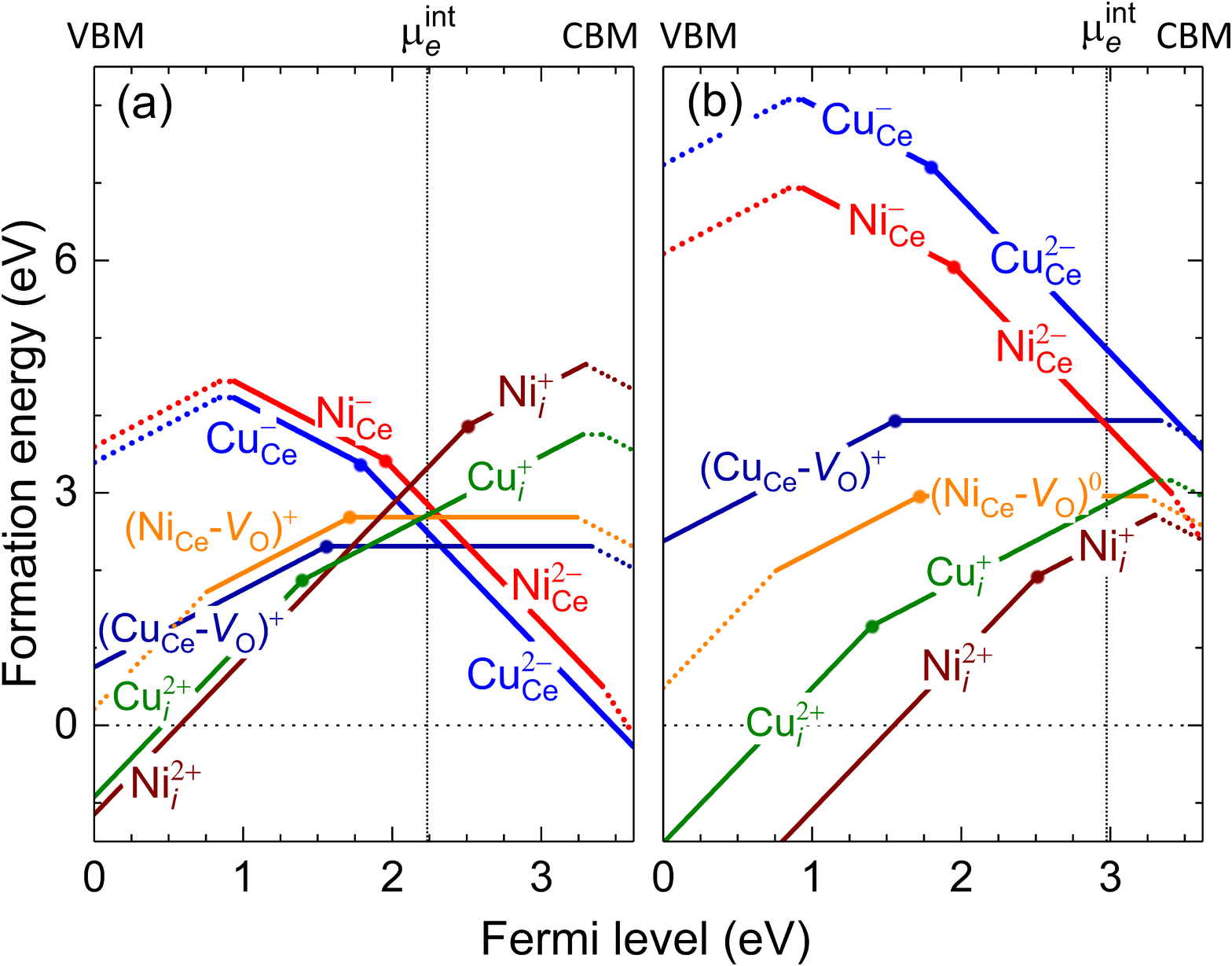}
\vspace{-0.5cm}
\caption{Formation energies of Cu- and Ni-related defects in CeO$_2$ obtained under (a) condition {\bf A} and (b) condition {\bf B}.}
\label{fig;fe;cuni}
\end{figure}

\begin{figure}
\centering
\includegraphics[width=\linewidth]{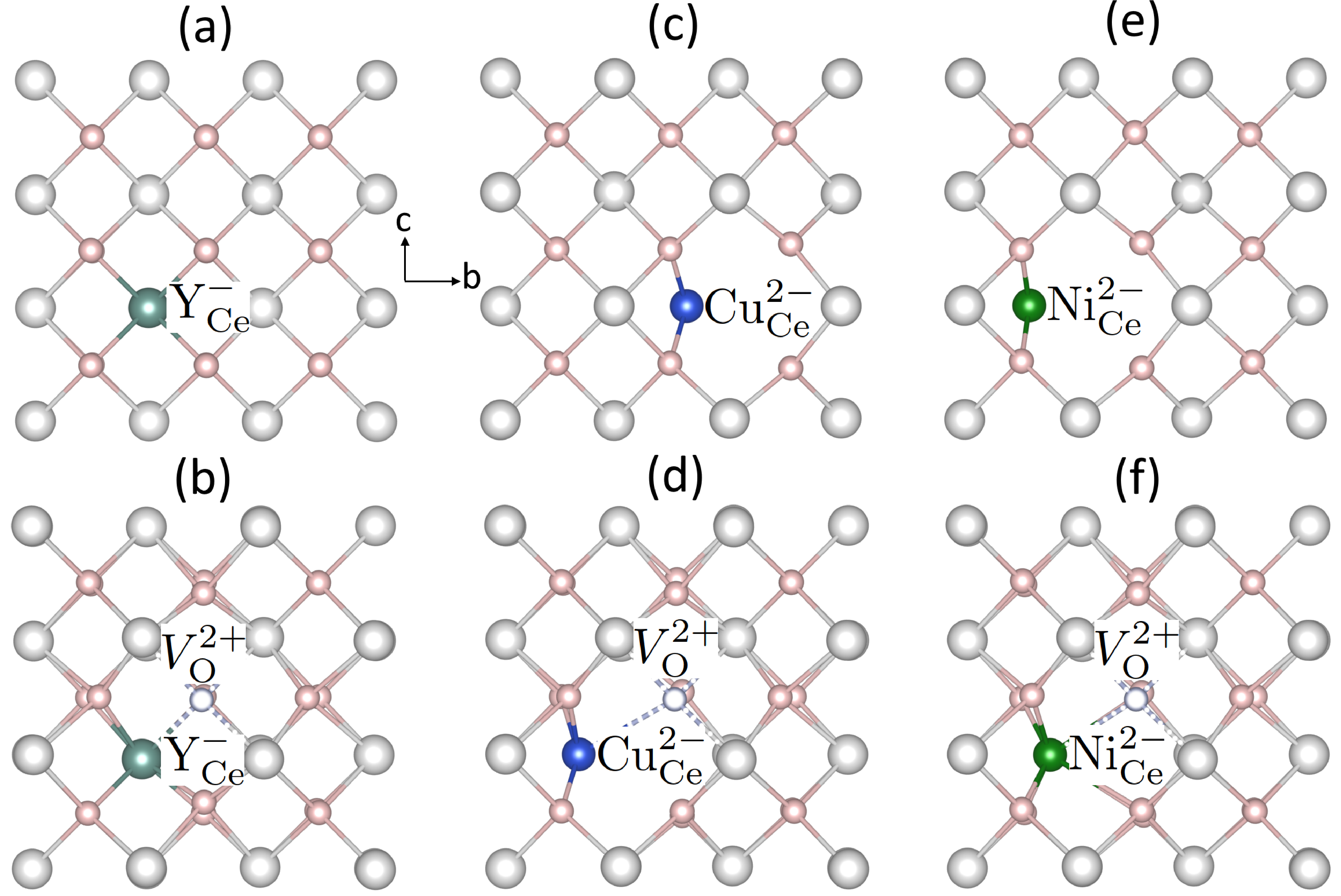}
\vspace{-0.5cm}
\caption{Structures of representative metal dopants and related defect complexes in CeO$_2$: (a) Y$_{\rm Ce}^-$, (b) (Y$_{\rm Ce}$-$V_{\rm O}$)$^+$, (c) Cu$_{\rm Ce}^{2-}$, (d) (Cu$_{\rm Ce}$-$V_{\rm O}$)$^0$, (e) Ni$_{\rm Ce}^{2-}$, and (f) (Ni$_{\rm Ce}$-$V_{\rm O}$)$^0$.}
\label{fig;struct;metals}
\end{figure}

Figures~\ref{fig;fe;y} and \ref{fig;fe;cuni} show the calculated formation energies of select metal dopants in CeO$_2$. In the following discussion, we focus on the stable charge states (ionic defect configurations) presented by the solid energy lines in the formation-energy plot as other (nominal) charge states (dotted energy lines) are just complexes of the stable ionic defects and $\eta_{\rm Ce}^-$ or $\eta_{\rm O}^+$ (or $\eta_{\rm O}^\ast$ in the case of Y$_{\rm Ce}$, where $\eta_{\rm O}^\ast$ is an electron hole delocalized over two O sites).

We find that the substitutional Y$_{\rm Ce}$ is stable as Y$_{\rm Ce}^-$ in which Y$^{3+}$ replaces Ce$^{4+}$ at a Ce site. The Y--O distance is 2.36 {\AA}, comparable to the Ce--O bond length (2.33 {\AA}) in the perfect bulk material; see Fig.~\ref{fig;struct;metals}(a). The small lattice distortion is due to the small difference in the ionic radii of eight-fold coordinated Ce$^{4+}$ (0.97 {\AA}) and Y$^{3+}$ (1.02 {\AA}) \cite{Shannon1976}. Due to the Coulombic attraction, Y$_{\rm Ce}^-$ and $V_{\rm O}^{2+}$ can come close and form (Y$_{\rm Ce}$-$V_{\rm O}$)$^+$ with the binding energy of 0.44 eV; see Fig.~\ref{fig;struct;metals}(b). This complex can capture an $\eta_{\rm Ce}^-$ to become (Y$_{\rm Ce}$-$V_{\rm O}$)$^0$ with a binding energy of 0.59 eV, or combine with another Y$_{\rm Ce}^-$ to form a (2Y$_{\rm Ce}$-$V_{\rm O}$)$^0$ with a binding energy of 0.73 eV. (2Y$_{\rm Ce}$-$V_{\rm O}$)$^0$ is found to be lower in energy than (Y$_{\rm Ce}$-$V_{\rm O}$)$^0$, except in highly reducing environments (e.g., under condition {\bf B}).  

The interstitial Y$_i$ is stable as Y$_i^{3+}$, which is an extra Y$^{3+}$ ion at the center of a Ce octahedron. Y$_i^{3+}$ is always in much higher in energy than Y$_{\rm Ce}^-$ and related defect complexes. This is due to the large ionic radius of Y$^{3+}$. 

Next, Cu$_{\rm Ce}$ is stable as Cu$_{\rm Ce}^{2-}$ (i.e., Cu$^{2+}$ $3d^9$, $S=1/2$, at the Ce$^{4+}$ site) and/or Cu$_{\rm Ce}^-$ (i.e., Cu$^{3+}$ $3d^8$, $S=0$, at the Ce$^{4+}$ site). The Cu$^+$ ion is thus not stable at the substitutional site. Cu$_{\rm Ce}$ is significantly off-center, by 0.99 {\AA} (1.11 {\AA}) in the case of Cu$_{\rm Ce}^{2-}$ (Cu$_{\rm Ce}^{-}$), and forms a (slightly square pyramidal distorted) square planar with four oxygens; see Fig.~\ref{fig;struct;metals}(c). The large lattice distortion is consistent with the large difference in the ionic radii of Cu$^{2+}$ (Cu$^{3+}$) and Ce$^{4+}$ \cite{Shannon1976}. Possible Cu-related defect complexes include (Cu$_{\rm Ce}$-$V_{\rm O}$)$^0$, a complex of Cu$_{\rm Ce}^{2-}$ and $V_{\rm O}^{2+}$ with a binding energy of 1.35 eV, see Fig.~\ref{fig;struct;metals}(d), and (Cu$_{\rm Ce}$-$V_{\rm O}$)$^+$, a complex of Cu$_{\rm Ce}^{-}$ and $V_{\rm O}^{2+}$ with a binding energy of 1.11 eV. Note that our results for Cu are different from those of Sun and Yildiz~\cite{Sun2019JPCC}; e.g., (Cu$_{\rm Ce}$-$V_{\rm O}$)$^+$ does not appear in their formation-energy plot.

The interstitial Cu$_i$ is electronically stable as Cu$_i^{2+}$ (i.e, Cu$^{2+}$ $3d^9$, $S=1/2$) or Cu$_i^{+}$ (Cu$^{+}$ $3d^{10}$, $S=0$), with the Cu$^{2+}$ (or Cu$^{+}$) ion at the center of the Ce octahedron. We find that, at the Fermi-level position $\mu_e^{\rm int}$, the formation energy of Cu$_i^{+}$ is higher than Cu$_{\rm Ce}^{2-}$ for $0$ eV $\leq \mu_{\rm O} \lessapprox -1.10$ eV [which includes condition {\bf A}--see Fig.~\ref{fig;fe;cuni}(a)--and condition {\bf C}] and lower for $-1.10$ eV $\lessapprox \mu_{\rm O} \leq -3.09$ eV [which includes condition {\bf B}, see Fig.~\ref{fig;fe;cuni}(b)]. Cu$_i^{+}$ and Cu$_{\rm Ce}^{2-}$ have comparable formation energies for $\mu_{\rm O} \approx -1.10 eV$ [i.e., near condition {\bf A}, see Fig.~\ref{fig;fe;cuni}(a)]. Our results suggest that under normal conditions, Cu may exist both as Cu$^{2+}$ at the substitutional site and as Cu$^{+}$ at the interstitial site. Under reducing conditions, on the other hand, Cu$_i^{+}$ (i.e., Cu$^{+}$) becomes the dominant Cu species; see, e.g., Fig.~\ref{fig;fe;cuni}(b). This appears to be consistent with experimental observations of multiple Cu oxidation states in Cu-doped CeO$_2$~\cite{Sartoretti2023CT,DavoQuinonero2020ACSC}.

Finally, Ni$_{\rm Ce}$ is stable as Ni$_{\rm Ce}^-$ (i.e., low-spin Ni$^{3+}$ $3d^7$, $S=1/2$, at the Ce$^{4+}$ site) and Ni$_{\rm Ce}^{2-}$ (i.e., low-spin Ni$^{2+}$ $3d^8$, $S=0$, at the Ce$^{4+}$ site). Like Cu, Ni is significantly off-center, by 1.07 {\AA} (1.17 {\AA}) in the case of Ni$_{\rm Ce}^{-}$ (Ni$_{\rm Ce}^{2-}$), and forms a (slightly square pyramidal distorted) square planar with four oxygen atoms; see Fig.~\ref{fig;struct;metals}(e). The high-spin state of Ni$^{2+}$ ($S=2$) is only 5 meV higher in energy than the low-spin one; i.e., they are essentially degenerate. Defect complexes with oxygen vacancies include (Ni$_{\rm Ce}$-$V_{\rm O}$)$^0$, a complex of Ni$_{\rm Ce}^{2-}$ and $V_{\rm O}^{2+}$ with a binding energy of 1.33 eV, see Fig.~\ref{fig;struct;metals}(f), and (Ni$_{\rm Ce}$-$V_{\rm O}$)$^+$, a complex of Ni$_{\rm Ce}^{-}$ and $V_{\rm O}^{2+}$ with a binding energy of 1.11 eV. Similar to Cu$_i$, the interstitial Ni$_i$ is also stable electronically as Ni$_i^{2+}$ (where the low-spin and high-spin states Ni$^{2+}$ are degenerate in energy) or Ni$_i^+$ (i.e., Ni$^+$ $3d^9$; $S=1/2$). At $\mu_e^{\rm int}$, Ni$_i^+$ is higher in energy than Ni$_{\rm Ce}^{2-}$ under conditions {\bf A} and {\bf C}, but lower under condition {\bf B}; see Fig.~\ref{fig;fe;cuni}(b); the crossover point is at $\mu_{\rm O} \approx -1.59$ eV. Note that Wang et al.~\cite{Wang2010JPCC} also found the interstitial Ni to be stable as Ni$^+$ in their DFT$+$$U$ calculations.   

Overall, we find that, under experimentally relevant conditions such as in Fig.~\ref{fig;fe;y} and Fig.~\ref{fig;fe;cuni}, the substitutional dopants Y, Cu, and Ni in CeO$_2$ are stable, respectively, as Y$_{\rm Ce}^-$, Cu$_{\rm Ce}^{2-}$, and Ni$_{\rm Ce}^{2-}$ (either as unassociated defects or in complexes with oxygen vacancies) at the Fermi level $\mu_e^{\rm int}$ determined by native defects. The Y, Cu, and Ni interstitials, on the other hand, are stable as Y$_i^{3+}$, Cu$_i^+$, and Ni$_i^{2+}$, respectively. For Cu and Ni, the lattice site preference (substitutional vs.~interstitial) and the charge and spin state of the metal dopants depends on actual experimental conditions. The difference in the structure and energetics between Cu (or Ni) versus Y can be ascribed to the Shannon ionic radius of Cu (Ni) being smaller than that of Y~\cite{Shannon1976}. As negatively charged (acceptor-like) defects, Y$_{\rm Ce}^-$, Cu$_{\rm Ce}^{2-}$, and Ni$_{\rm Ce}^{2-}$ can facilitate the formation of charge-compensating, positively charged oxygen vacancies, without increasing the concentration of Ce$^{3+}$ and thus without increasing the electronic conductivity. As positively charged (donor-like) defects and if present with a large concentration, Y$_i^{3+}$, Cu$_i^+$, and Ni$_i^{2+}$ can lower the formation energy of $\eta_{\rm Ce}^-$ and thus increase the Ce$^{3+}$ concentration. See, e.g., Ref.~\citenum{Hoang2018JPCM} for a further discussion of effects of doping in complex metal oxides.

As a general note, having a finite binding energy does not mean a complex will readily form. Under thermal equilibrium, the binding energy needs to be greater than the larger of the formation energies of the individual defects for the complex to have higher concentration than its constituents \cite{walle:3851}. Most of the relevant defect complexes considered in this work have a small binding energy. This suggests that their concentration is expected to be much smaller than that of the isolated constituents. 

Experimentally, Thurber et al.~\cite{Thurber2007PRB} reported x-ray photoelectron spectroscopy studies of Ce$_{1-x}$Ni$_x$O$_{2}$ nanoparticles ($0.01\leq x \leq$ 0.20) which showed that nickel is substitutionally incorporated as Ni$^{2+}$ at the Ce site in the bulk; there is also contribution from Ni$^{3+}$ surface states. Derafa et al.~\cite{Derafa2018RSCA}, on the other hand, reported the incorporation of nickel into the lattice in the form of Ni$^{3+}$ species in Ni$_{0.1}$Ce$_{0.9}$O$_{2-x}$ nanoparticles. The determination of the identity of Ni species in CeO$_2$ is expected to be challenging as, according to our results, the dopant can be incorporated into the lattice at different lattice sites and exist in different charge and spin states. Regarding the effects of substitutional metal doping, experimental studies appear to be unanimous on the role of Y, Cu, and Ni in promoting oxygen vacancies \cite{Zhang2019CC,Ranjith2018SCE,Polychronopoulou2021AMI,Thurber2007PRB,Derafa2018RSCA,Fuda1984JPCS,Wang1981SSI,Lee2014JPCC}, which is consistent with our above analysis.

\section{Conclusions}

We have carried out a hybrid density-functional study of defect physics in bulk CeO$_2$. The (negatively charged) electron polaron ($\eta_{\rm Ce}^-$, i.e., Ce$^{3+}$) and the positively charged oxygen vacancy ($V_{\rm O}^{2+}$) are found to be the dominant native defects under conditions ranging from extreme oxidizing to highly reducing. The migration barriers of and the binding energy between $\eta_{\rm Ce}^-$ and $V_{\rm O}^{2+}$ are in good agreement with experiments. Hydrogen is stable either in the hydroxyl (H$_i^+$) or hydride (H$_{\rm O}^+$) configuration, depending actual experimental conditions. The interstitial H$_i^{+}$ is highly mobile in the bulk. Yttrium is most stable as the substitutional lattice site; the lattice preference (substitutional versus interstitial) of copper and nickel, on the other hand, depends on actual conditions during materials preparation, and the dopants can exist in different charge and spin states. In light of the results, we discuss the effects of metal doping on the formation of electron polarons and oxygen vacancies. 

This work thus provides guidance for experimental defect characterization and defect-controlled synthesis. It also serves as a methodological template and a basis for further investigation of other dopants and impurities. A detailed understanding of the synthesis--(defect) structure--property relationship such as that developed here would ultimately lead to the design and discovery of CeO$_2$-based materials with better performance.   

\begin{acknowledgments}

M.D.J.~was supported by the U.S.~Office of Naval Research through the U.S.~Naval Research Laboratory's core fundamental research program. This work used resources of the Center for Computationally Assisted Science and Technology (CCAST) at North Dakota State University, which were made possible in part by National Science Foundation Major Research Instrumentation (MRI) Award No.~2019077.

\end{acknowledgments}


\begin{thebibliography}{68}%
\makeatletter
\providecommand \@ifxundefined [1]{%
 \@ifx{#1\undefined}
}%
\providecommand \@ifnum [1]{%
 \ifnum #1\expandafter \@firstoftwo
 \else \expandafter \@secondoftwo
 \fi
}%
\providecommand \@ifx [1]{%
 \ifx #1\expandafter \@firstoftwo
 \else \expandafter \@secondoftwo
 \fi
}%
\providecommand \natexlab [1]{#1}%
\providecommand \enquote  [1]{``#1''}%
\providecommand \bibnamefont  [1]{#1}%
\providecommand \bibfnamefont [1]{#1}%
\providecommand \citenamefont [1]{#1}%
\providecommand \href@noop [0]{\@secondoftwo}%
\providecommand \href [0]{\begingroup \@sanitize@url \@href}%
\providecommand \@href[1]{\@@startlink{#1}\@@href}%
\providecommand \@@href[1]{\endgroup#1\@@endlink}%
\providecommand \@sanitize@url [0]{\catcode `\\12\catcode `\$12\catcode
  `\&12\catcode `\#12\catcode `\^12\catcode `\_12\catcode `\%12\relax}%
\providecommand \@@startlink[1]{}%
\providecommand \@@endlink[0]{}%
\providecommand \url  [0]{\begingroup\@sanitize@url \@url }%
\providecommand \@url [1]{\endgroup\@href {#1}{\urlprefix }}%
\providecommand \urlprefix  [0]{URL }%
\providecommand \Eprint [0]{\href }%
\providecommand \doibase [0]{https://doi.org/}%
\providecommand \selectlanguage [0]{\@gobble}%
\providecommand \bibinfo  [0]{\@secondoftwo}%
\providecommand \bibfield  [0]{\@secondoftwo}%
\providecommand \translation [1]{[#1]}%
\providecommand \BibitemOpen [0]{}%
\providecommand \bibitemStop [0]{}%
\providecommand \bibitemNoStop [0]{.\EOS\space}%
\providecommand \EOS [0]{\spacefactor3000\relax}%
\providecommand \BibitemShut  [1]{\csname bibitem#1\endcsname}%
\let\auto@bib@innerbib\@empty
\bibitem [{\citenamefont {Montini}\ \emph {et~al.}(2016)\citenamefont
  {Montini}, \citenamefont {Melchionna}, \citenamefont {Monai},\ and\
  \citenamefont {Fornasiero}}]{Montini2016CR}%
  \BibitemOpen
  \bibfield  {author} {\bibinfo {author} {\bibfnamefont {T.}~\bibnamefont
  {Montini}}, \bibinfo {author} {\bibfnamefont {M.}~\bibnamefont {Melchionna}},
  \bibinfo {author} {\bibfnamefont {M.}~\bibnamefont {Monai}},\ and\ \bibinfo
  {author} {\bibfnamefont {P.}~\bibnamefont {Fornasiero}},\ }\bibfield  {title}
  {\bibinfo {title} {{Fundamentals and Catalytic Applications of CeO$_2$-Based
  Materials}},\ }\href {https://doi.org/10.1021/acs.chemrev.5b00603} {\bibfield
   {journal} {\bibinfo  {journal} {Chem. Rev.}\ }\textbf {\bibinfo {volume}
  {116}},\ \bibinfo {pages} {5987} (\bibinfo {year} {2016})}\BibitemShut
  {NoStop}%
\bibitem [{\citenamefont {Hoang}\ and\ \citenamefont
  {Johannes}(2018)}]{Hoang2018JPCM}%
  \BibitemOpen
  \bibfield  {author} {\bibinfo {author} {\bibfnamefont {K.}~\bibnamefont
  {Hoang}}\ and\ \bibinfo {author} {\bibfnamefont {M.~D.}\ \bibnamefont
  {Johannes}},\ }\bibfield  {title} {\bibinfo {title} {Defect physics in
  complex energy materials},\ }\href
  {http://stacks.iop.org/0953-8984/30/i=29/a=293001} {\bibfield  {journal}
  {\bibinfo  {journal} {J. Phys.: Condens. Matter}\ }\textbf {\bibinfo {volume}
  {30}},\ \bibinfo {pages} {293001} (\bibinfo {year} {2018})}\BibitemShut
  {NoStop}%
\bibitem [{\citenamefont {Paier}\ \emph {et~al.}(2013)\citenamefont {Paier},
  \citenamefont {Penschke},\ and\ \citenamefont {Sauer}}]{Paier2013CR}%
  \BibitemOpen
  \bibfield  {author} {\bibinfo {author} {\bibfnamefont {J.}~\bibnamefont
  {Paier}}, \bibinfo {author} {\bibfnamefont {C.}~\bibnamefont {Penschke}},\
  and\ \bibinfo {author} {\bibfnamefont {J.}~\bibnamefont {Sauer}},\ }\bibfield
   {title} {\bibinfo {title} {{Oxygen Defects and Surface Chemistry of Ceria:
  Quantum Chemical Studies Compared to Experiment}},\ }\href
  {https://doi.org/10.1021/cr3004949} {\bibfield  {journal} {\bibinfo
  {journal} {Chem. Rev.}\ }\textbf {\bibinfo {volume} {113}},\ \bibinfo {pages}
  {3949} (\bibinfo {year} {2013})}\BibitemShut {NoStop}%
\bibitem [{\citenamefont {Schmitt}\ \emph {et~al.}(2020)\citenamefont
  {Schmitt}, \citenamefont {Nenning}, \citenamefont {Kraynis}, \citenamefont
  {Korobko}, \citenamefont {Frenkel}, \citenamefont {Lubomirsky}, \citenamefont
  {Haile},\ and\ \citenamefont {Rupp}}]{Schmitt2020CSR}%
  \BibitemOpen
  \bibfield  {author} {\bibinfo {author} {\bibfnamefont {R.}~\bibnamefont
  {Schmitt}}, \bibinfo {author} {\bibfnamefont {A.}~\bibnamefont {Nenning}},
  \bibinfo {author} {\bibfnamefont {O.}~\bibnamefont {Kraynis}}, \bibinfo
  {author} {\bibfnamefont {R.}~\bibnamefont {Korobko}}, \bibinfo {author}
  {\bibfnamefont {A.~I.}\ \bibnamefont {Frenkel}}, \bibinfo {author}
  {\bibfnamefont {I.}~\bibnamefont {Lubomirsky}}, \bibinfo {author}
  {\bibfnamefont {S.~M.}\ \bibnamefont {Haile}},\ and\ \bibinfo {author}
  {\bibfnamefont {J.~L.~M.}\ \bibnamefont {Rupp}},\ }\bibfield  {title}
  {\bibinfo {title} {A review of defect structure and chemistry in ceria and
  its solid solutions},\ }\href {https://doi.org/10.1039/C9CS00588A} {\bibfield
   {journal} {\bibinfo  {journal} {Chem. Soc. Rev.}\ }\textbf {\bibinfo
  {volume} {49}},\ \bibinfo {pages} {554} (\bibinfo {year} {2020})}\BibitemShut
  {NoStop}%
\bibitem [{\citenamefont {Shoko}\ \emph {et~al.}(2011)\citenamefont {Shoko},
  \citenamefont {Smith},\ and\ \citenamefont {McKenzie}}]{Shoko2011JPCS}%
  \BibitemOpen
  \bibfield  {author} {\bibinfo {author} {\bibfnamefont {E.}~\bibnamefont
  {Shoko}}, \bibinfo {author} {\bibfnamefont {M.}~\bibnamefont {Smith}},\ and\
  \bibinfo {author} {\bibfnamefont {R.~H.}\ \bibnamefont {McKenzie}},\
  }\bibfield  {title} {\bibinfo {title} {Charge distribution and transport
  properties in reduced ceria phases: {A} review},\ }\href
  {https://doi.org/10.1016/j.jpcs.2011.09.002} {\bibfield  {journal} {\bibinfo
  {journal} {J. Phys. Chem. Solids}\ }\textbf {\bibinfo {volume} {72}},\
  \bibinfo {pages} {1482} (\bibinfo {year} {2011})}\BibitemShut {NoStop}%
\bibitem [{\citenamefont {Blumenthal}\ and\ \citenamefont
  {Hofmaier}(1974)}]{Blumenthal1974JES}%
  \BibitemOpen
  \bibfield  {author} {\bibinfo {author} {\bibfnamefont {R.~N.}\ \bibnamefont
  {Blumenthal}}\ and\ \bibinfo {author} {\bibfnamefont {R.~L.}\ \bibnamefont
  {Hofmaier}},\ }\bibfield  {title} {\bibinfo {title} {{The Temperature and
  Compositional Dependence of the Electrical Conductivity of Nonstoichiometric
  CeO$_{2-x}$}},\ }\href {https://doi.org/10.1149/1.2396805} {\bibfield
  {journal} {\bibinfo  {journal} {J. Electrochem. Soc.}\ }\textbf {\bibinfo
  {volume} {121}},\ \bibinfo {pages} {126} (\bibinfo {year}
  {1974})}\BibitemShut {NoStop}%
\bibitem [{\citenamefont {Tuller}\ and\ \citenamefont
  {Nowick}(1977)}]{Tuller1977JPCS}%
  \BibitemOpen
  \bibfield  {author} {\bibinfo {author} {\bibfnamefont {H.}~\bibnamefont
  {Tuller}}\ and\ \bibinfo {author} {\bibfnamefont {A.}~\bibnamefont
  {Nowick}},\ }\bibfield  {title} {\bibinfo {title} {Small polaron electron
  transport in reduced {CeO$_2$} single crystals},\ }\href
  {https://doi.org/10.1016/0022-3697(77)90124-X} {\bibfield  {journal}
  {\bibinfo  {journal} {J. Phys. Chem. Solids}\ }\textbf {\bibinfo {volume}
  {38}},\ \bibinfo {pages} {859} (\bibinfo {year} {1977})}\BibitemShut
  {NoStop}%
\bibitem [{\citenamefont {Naik}\ and\ \citenamefont
  {Tien}(1978)}]{Naik1978JPCS}%
  \BibitemOpen
  \bibfield  {author} {\bibinfo {author} {\bibfnamefont {I.}~\bibnamefont
  {Naik}}\ and\ \bibinfo {author} {\bibfnamefont {T.}~\bibnamefont {Tien}},\
  }\bibfield  {title} {\bibinfo {title} {Small-polaron mobility in
  nonstoichiometric cerium dioxide},\ }\href
  {https://doi.org/10.1016/0022-3697(78)90059-8} {\bibfield  {journal}
  {\bibinfo  {journal} {J. Phys. Chem. Solids}\ }\textbf {\bibinfo {volume}
  {39}},\ \bibinfo {pages} {311} (\bibinfo {year} {1978})}\BibitemShut
  {NoStop}%
\bibitem [{\citenamefont {Lee}\ \emph {et~al.}(2022)\citenamefont {Lee},
  \citenamefont {Kim}, \citenamefont {Sun}, \citenamefont {Lee}, \citenamefont
  {Kwon}, \citenamefont {Hwang}, \citenamefont {Seo}, \citenamefont {Paik},\
  and\ \citenamefont {Song}}]{Lee2022NJC}%
  \BibitemOpen
  \bibfield  {author} {\bibinfo {author} {\bibfnamefont {K.}~\bibnamefont
  {Lee}}, \bibinfo {author} {\bibfnamefont {S.}~\bibnamefont {Kim}}, \bibinfo
  {author} {\bibfnamefont {S.}~\bibnamefont {Sun}}, \bibinfo {author}
  {\bibfnamefont {G.}~\bibnamefont {Lee}}, \bibinfo {author} {\bibfnamefont
  {J.}~\bibnamefont {Kwon}}, \bibinfo {author} {\bibfnamefont {J.}~\bibnamefont
  {Hwang}}, \bibinfo {author} {\bibfnamefont {J.}~\bibnamefont {Seo}}, \bibinfo
  {author} {\bibfnamefont {U.}~\bibnamefont {Paik}},\ and\ \bibinfo {author}
  {\bibfnamefont {T.}~\bibnamefont {Song}},\ }\bibfield  {title} {\bibinfo
  {title} {Hydrogenated ceria nanoparticles for high-efficiency silicate
  adsorption},\ }\href {https://doi.org/10.1039/D2NJ04043C} {\bibfield
  {journal} {\bibinfo  {journal} {New J. Chem.}\ }\textbf {\bibinfo {volume}
  {46}},\ \bibinfo {pages} {20572} (\bibinfo {year} {2022})}\BibitemShut
  {NoStop}%
\bibitem [{\citenamefont {Fierro}\ \emph {et~al.}(1987)\citenamefont {Fierro},
  \citenamefont {Soria}, \citenamefont {Sanz},\ and\ \citenamefont
  {Rojo}}]{Fierro1987JSSC}%
  \BibitemOpen
  \bibfield  {author} {\bibinfo {author} {\bibfnamefont {J.}~\bibnamefont
  {Fierro}}, \bibinfo {author} {\bibfnamefont {J.}~\bibnamefont {Soria}},
  \bibinfo {author} {\bibfnamefont {J.}~\bibnamefont {Sanz}},\ and\ \bibinfo
  {author} {\bibfnamefont {J.}~\bibnamefont {Rojo}},\ }\bibfield  {title}
  {\bibinfo {title} {Induced changes in ceria by thermal treatments under
  vacuum or hydrogen},\ }\href {https://doi.org/10.1016/0022-4596(87)90230-1}
  {\bibfield  {journal} {\bibinfo  {journal} {J. Solid State Chem.}\ }\textbf
  {\bibinfo {volume} {66}},\ \bibinfo {pages} {154} (\bibinfo {year}
  {1987})}\BibitemShut {NoStop}%
\bibitem [{\citenamefont {Bruce}\ \emph {et~al.}(1996)\citenamefont {Bruce},
  \citenamefont {Hoang}, \citenamefont {Hughes},\ and\ \citenamefont
  {Turney}}]{Bruce1996ACA}%
  \BibitemOpen
  \bibfield  {author} {\bibinfo {author} {\bibfnamefont {L.~A.}\ \bibnamefont
  {Bruce}}, \bibinfo {author} {\bibfnamefont {M.}~\bibnamefont {Hoang}},
  \bibinfo {author} {\bibfnamefont {A.~E.}\ \bibnamefont {Hughes}},\ and\
  \bibinfo {author} {\bibfnamefont {T.~W.}\ \bibnamefont {Turney}},\ }\bibfield
   {title} {\bibinfo {title} {Surface area control during the synthesis and
  reduction of high area ceria catalyst supports},\ }\href
  {https://doi.org/10.1016/0926-860X(95)00217-0} {\bibfield  {journal}
  {\bibinfo  {journal} {Appl. Catal. A: Gen.}\ }\textbf {\bibinfo {volume}
  {134}},\ \bibinfo {pages} {351} (\bibinfo {year} {1996})}\BibitemShut
  {NoStop}%
\bibitem [{\citenamefont {Mao}\ \emph {et~al.}(2024)\citenamefont {Mao},
  \citenamefont {Gong}, \citenamefont {Gu}, \citenamefont {Wilde},
  \citenamefont {Chen}, \citenamefont {Fukutani}, \citenamefont {Matsuzaki},
  \citenamefont {Fugetsu}, \citenamefont {Sakata},\ and\ \citenamefont
  {Terai}}]{Mao2024IJHE}%
  \BibitemOpen
  \bibfield  {author} {\bibinfo {author} {\bibfnamefont {W.}~\bibnamefont
  {Mao}}, \bibinfo {author} {\bibfnamefont {W.}~\bibnamefont {Gong}}, \bibinfo
  {author} {\bibfnamefont {Z.}~\bibnamefont {Gu}}, \bibinfo {author}
  {\bibfnamefont {M.}~\bibnamefont {Wilde}}, \bibinfo {author} {\bibfnamefont
  {J.}~\bibnamefont {Chen}}, \bibinfo {author} {\bibfnamefont {K.}~\bibnamefont
  {Fukutani}}, \bibinfo {author} {\bibfnamefont {H.}~\bibnamefont {Matsuzaki}},
  \bibinfo {author} {\bibfnamefont {B.}~\bibnamefont {Fugetsu}}, \bibinfo
  {author} {\bibfnamefont {I.}~\bibnamefont {Sakata}},\ and\ \bibinfo {author}
  {\bibfnamefont {T.}~\bibnamefont {Terai}},\ }\bibfield  {title} {\bibinfo
  {title} {Hydrogen diffusion in cerium oxide thin films fabricated by pulsed
  laser deposition},\ }\href {https://doi.org/10.1016/j.ijhydene.2023.08.264}
  {\bibfield  {journal} {\bibinfo  {journal} {Int. J. Hydrogen Energy}\
  }\textbf {\bibinfo {volume} {50}},\ \bibinfo {pages} {969} (\bibinfo {year}
  {2024})}\BibitemShut {NoStop}%
\bibitem [{\citenamefont {Zhang}\ \emph {et~al.}(2019)\citenamefont {Zhang},
  \citenamefont {Zhao}, \citenamefont {Liu}, \citenamefont {Li}, \citenamefont
  {Wang}, \citenamefont {Wang}, \citenamefont {Zhang}, \citenamefont {Zhang},\
  and\ \citenamefont {Zhao}}]{Zhang2019CC}%
  \BibitemOpen
  \bibfield  {author} {\bibinfo {author} {\bibfnamefont {S.}~\bibnamefont
  {Zhang}}, \bibinfo {author} {\bibfnamefont {C.}~\bibnamefont {Zhao}},
  \bibinfo {author} {\bibfnamefont {Y.}~\bibnamefont {Liu}}, \bibinfo {author}
  {\bibfnamefont {W.}~\bibnamefont {Li}}, \bibinfo {author} {\bibfnamefont
  {J.}~\bibnamefont {Wang}}, \bibinfo {author} {\bibfnamefont {G.}~\bibnamefont
  {Wang}}, \bibinfo {author} {\bibfnamefont {Y.}~\bibnamefont {Zhang}},
  \bibinfo {author} {\bibfnamefont {H.}~\bibnamefont {Zhang}},\ and\ \bibinfo
  {author} {\bibfnamefont {H.}~\bibnamefont {Zhao}},\ }\bibfield  {title}
  {\bibinfo {title} {{Cu doping in CeO$_2$ to form multiple oxygen vacancies
  for dramatically enhanced ambient N$_2$ reduction performance}},\ }\href
  {https://doi.org/10.1039/C9CC00123A} {\bibfield  {journal} {\bibinfo
  {journal} {Chem. Commun.}\ }\textbf {\bibinfo {volume} {55}},\ \bibinfo
  {pages} {2952} (\bibinfo {year} {2019})}\BibitemShut {NoStop}%
\bibitem [{\citenamefont {Ranjith}\ \emph {et~al.}(2018)\citenamefont
  {Ranjith}, \citenamefont {Dong}, \citenamefont {Lu}, \citenamefont {Huang},
  \citenamefont {Chen}, \citenamefont {Saravanan}, \citenamefont {Asokan},\
  and\ \citenamefont {Rajendra~Kumar}}]{Ranjith2018SCE}%
  \BibitemOpen
  \bibfield  {author} {\bibinfo {author} {\bibfnamefont {K.~S.}\ \bibnamefont
  {Ranjith}}, \bibinfo {author} {\bibfnamefont {C.-L.}\ \bibnamefont {Dong}},
  \bibinfo {author} {\bibfnamefont {Y.-R.}\ \bibnamefont {Lu}}, \bibinfo
  {author} {\bibfnamefont {Y.-C.}\ \bibnamefont {Huang}}, \bibinfo {author}
  {\bibfnamefont {C.-L.}\ \bibnamefont {Chen}}, \bibinfo {author}
  {\bibfnamefont {P.}~\bibnamefont {Saravanan}}, \bibinfo {author}
  {\bibfnamefont {K.}~\bibnamefont {Asokan}},\ and\ \bibinfo {author}
  {\bibfnamefont {R.~T.}\ \bibnamefont {Rajendra~Kumar}},\ }\bibfield  {title}
  {\bibinfo {title} {{Evolution of Visible Photocatalytic Properties of
  Cu-Doped CeO$_2$ Nanoparticles: Role of Cu$^{2+}$-Mediated Oxygen Vacancies
  and the Mixed-Valence States of Ce Ions}},\ }\href
  {https://doi.org/10.1021/acssuschemeng.8b00848} {\bibfield  {journal}
  {\bibinfo  {journal} {ACS Sustain. Chem. Eng.}\ }\textbf {\bibinfo {volume}
  {6}},\ \bibinfo {pages} {8536} (\bibinfo {year} {2018})}\BibitemShut
  {NoStop}%
\bibitem [{\citenamefont {Sartoretti}\ \emph {et~al.}(2023)\citenamefont
  {Sartoretti}, \citenamefont {Novara}, \citenamefont {Paganini}, \citenamefont
  {Chiesa}, \citenamefont {Castellino}, \citenamefont {Giorgis}, \citenamefont
  {Piumetti}, \citenamefont {Bensaid}, \citenamefont {Fino},\ and\
  \citenamefont {Russo}}]{Sartoretti2023CT}%
  \BibitemOpen
  \bibfield  {author} {\bibinfo {author} {\bibfnamefont {E.}~\bibnamefont
  {Sartoretti}}, \bibinfo {author} {\bibfnamefont {C.}~\bibnamefont {Novara}},
  \bibinfo {author} {\bibfnamefont {M.~C.}\ \bibnamefont {Paganini}}, \bibinfo
  {author} {\bibfnamefont {M.}~\bibnamefont {Chiesa}}, \bibinfo {author}
  {\bibfnamefont {M.}~\bibnamefont {Castellino}}, \bibinfo {author}
  {\bibfnamefont {F.}~\bibnamefont {Giorgis}}, \bibinfo {author} {\bibfnamefont
  {M.}~\bibnamefont {Piumetti}}, \bibinfo {author} {\bibfnamefont
  {S.}~\bibnamefont {Bensaid}}, \bibinfo {author} {\bibfnamefont
  {D.}~\bibnamefont {Fino}},\ and\ \bibinfo {author} {\bibfnamefont
  {N.}~\bibnamefont {Russo}},\ }\bibfield  {title} {\bibinfo {title}
  {{Investigation of Cu-doped ceria through a combined spectroscopic approach:
  Involvement of different catalytic sites in CO oxidation}},\ }\href
  {https://doi.org/10.1016/j.cattod.2023.02.014} {\bibfield  {journal}
  {\bibinfo  {journal} {Catal. Today}\ }\textbf {\bibinfo {volume} {420}},\
  \bibinfo {pages} {114037} (\bibinfo {year} {2023})}\BibitemShut {NoStop}%
\bibitem [{\citenamefont {Davó-Qui{\~n}onero}\ \emph
  {et~al.}(2020)\citenamefont {Davó-Qui{\~n}onero}, \citenamefont
  {Bailón-García}, \citenamefont {López-Rodríguez}, \citenamefont
  {Juan-Juan}, \citenamefont {Lozano-Castelló}, \citenamefont
  {García-Melchor}, \citenamefont {Herrera}, \citenamefont {Pellegrin},
  \citenamefont {Escudero},\ and\ \citenamefont
  {Bueno-López}}]{DavoQuinonero2020ACSC}%
  \BibitemOpen
  \bibfield  {author} {\bibinfo {author} {\bibfnamefont {A.}~\bibnamefont
  {Davó-Qui{\~n}onero}}, \bibinfo {author} {\bibfnamefont {E.}~\bibnamefont
  {Bailón-García}}, \bibinfo {author} {\bibfnamefont {S.}~\bibnamefont
  {López-Rodríguez}}, \bibinfo {author} {\bibfnamefont {J.}~\bibnamefont
  {Juan-Juan}}, \bibinfo {author} {\bibfnamefont {D.}~\bibnamefont
  {Lozano-Castelló}}, \bibinfo {author} {\bibfnamefont {M.}~\bibnamefont
  {García-Melchor}}, \bibinfo {author} {\bibfnamefont {F.~C.}\ \bibnamefont
  {Herrera}}, \bibinfo {author} {\bibfnamefont {E.}~\bibnamefont {Pellegrin}},
  \bibinfo {author} {\bibfnamefont {C.}~\bibnamefont {Escudero}},\ and\
  \bibinfo {author} {\bibfnamefont {A.}~\bibnamefont {Bueno-López}},\
  }\bibfield  {title} {\bibinfo {title} {{Insights into the Oxygen Vacancy
  Filling Mechanism in CuO/CeO$_2$ Catalysts: A Key Step Toward High
  Selectivity in Preferential CO Oxidation}},\ }\href
  {https://doi.org/10.1021/acscatal.0c00648} {\bibfield  {journal} {\bibinfo
  {journal} {ACS Catal.}\ }\textbf {\bibinfo {volume} {10}},\ \bibinfo {pages}
  {6532} (\bibinfo {year} {2020})}\BibitemShut {NoStop}%
\bibitem [{\citenamefont {Polychronopoulou}\ \emph {et~al.}(2021)\citenamefont
  {Polychronopoulou}, \citenamefont {AlKhoori}, \citenamefont {Efstathiou},
  \citenamefont {Jaoude}, \citenamefont {Damaskinos}, \citenamefont {Baker},
  \citenamefont {Almutawa}, \citenamefont {Anjum}, \citenamefont {Vasiliades},
  \citenamefont {Belabbes}, \citenamefont {Vega}, \citenamefont {Zedan},\ and\
  \citenamefont {Hinder}}]{Polychronopoulou2021AMI}%
  \BibitemOpen
  \bibfield  {author} {\bibinfo {author} {\bibfnamefont {K.}~\bibnamefont
  {Polychronopoulou}}, \bibinfo {author} {\bibfnamefont {A.~A.}\ \bibnamefont
  {AlKhoori}}, \bibinfo {author} {\bibfnamefont {A.~M.}\ \bibnamefont
  {Efstathiou}}, \bibinfo {author} {\bibfnamefont {M.~A.}\ \bibnamefont
  {Jaoude}}, \bibinfo {author} {\bibfnamefont {C.~M.}\ \bibnamefont
  {Damaskinos}}, \bibinfo {author} {\bibfnamefont {M.~A.}\ \bibnamefont
  {Baker}}, \bibinfo {author} {\bibfnamefont {A.}~\bibnamefont {Almutawa}},
  \bibinfo {author} {\bibfnamefont {D.~H.}\ \bibnamefont {Anjum}}, \bibinfo
  {author} {\bibfnamefont {M.~A.}\ \bibnamefont {Vasiliades}}, \bibinfo
  {author} {\bibfnamefont {A.}~\bibnamefont {Belabbes}}, \bibinfo {author}
  {\bibfnamefont {L.~F.}\ \bibnamefont {Vega}}, \bibinfo {author}
  {\bibfnamefont {A.~F.}\ \bibnamefont {Zedan}},\ and\ \bibinfo {author}
  {\bibfnamefont {S.~J.}\ \bibnamefont {Hinder}},\ }\bibfield  {title}
  {\bibinfo {title} {{Design Aspects of Doped CeO$_2$ for Low-Temperature
  Catalytic CO Oxidation: Transient Kinetics and DFT Approach}},\ }\href
  {https://doi.org/10.1021/acsami.1c02934} {\bibfield  {journal} {\bibinfo
  {journal} {ACS Appl. Mater. Interfaces}\ }\textbf {\bibinfo {volume} {13}},\
  \bibinfo {pages} {22391} (\bibinfo {year} {2021})}\BibitemShut {NoStop}%
\bibitem [{\citenamefont {Wrobel}\ \emph {et~al.}(1996)\citenamefont {Wrobel},
  \citenamefont {Lamonier}, \citenamefont {Bennani}, \citenamefont
  {D{'}Huysser},\ and\ \citenamefont {Aboukaïs}}]{Wrobel1996JCS}%
  \BibitemOpen
  \bibfield  {author} {\bibinfo {author} {\bibfnamefont {G.}~\bibnamefont
  {Wrobel}}, \bibinfo {author} {\bibfnamefont {C.}~\bibnamefont {Lamonier}},
  \bibinfo {author} {\bibfnamefont {A.}~\bibnamefont {Bennani}}, \bibinfo
  {author} {\bibfnamefont {A.}~\bibnamefont {D{'}Huysser}},\ and\ \bibinfo
  {author} {\bibfnamefont {A.}~\bibnamefont {Aboukaïs}},\ }\bibfield  {title}
  {\bibinfo {title} {{Effect of incorporation of copper or nickel on hydrogen
  storage in ceria. Mechanism of reduction}},\ }\href
  {https://doi.org/10.1039/FT9969202001} {\bibfield  {journal} {\bibinfo
  {journal} {J. Chem. Soc.{,} Faraday Trans.}\ }\textbf {\bibinfo {volume}
  {92}},\ \bibinfo {pages} {2001} (\bibinfo {year} {1996})}\BibitemShut
  {NoStop}%
\bibitem [{\citenamefont {Thurber}\ \emph {et~al.}(2007)\citenamefont
  {Thurber}, \citenamefont {Reddy}, \citenamefont {Shutthanandan},
  \citenamefont {Engelhard}, \citenamefont {Wang}, \citenamefont {Hays},\ and\
  \citenamefont {Punnoose}}]{Thurber2007PRB}%
  \BibitemOpen
  \bibfield  {author} {\bibinfo {author} {\bibfnamefont {A.}~\bibnamefont
  {Thurber}}, \bibinfo {author} {\bibfnamefont {K.~M.}\ \bibnamefont {Reddy}},
  \bibinfo {author} {\bibfnamefont {V.}~\bibnamefont {Shutthanandan}}, \bibinfo
  {author} {\bibfnamefont {M.~H.}\ \bibnamefont {Engelhard}}, \bibinfo {author}
  {\bibfnamefont {C.}~\bibnamefont {Wang}}, \bibinfo {author} {\bibfnamefont
  {J.}~\bibnamefont {Hays}},\ and\ \bibinfo {author} {\bibfnamefont
  {A.}~\bibnamefont {Punnoose}},\ }\bibfield  {title} {\bibinfo {title}
  {{Ferromagnetism in chemically synthesized $\mathrm{Ce}{\mathrm{O}}_{2}$
  nanoparticles by Ni doping}},\ }\href
  {https://doi.org/10.1103/PhysRevB.76.165206} {\bibfield  {journal} {\bibinfo
  {journal} {Phys. Rev. B}\ }\textbf {\bibinfo {volume} {76}},\ \bibinfo
  {pages} {165206} (\bibinfo {year} {2007})}\BibitemShut {NoStop}%
\bibitem [{\citenamefont {Derafa}\ \emph {et~al.}(2018)\citenamefont {Derafa},
  \citenamefont {Paloukis}, \citenamefont {Mewafy}, \citenamefont {Baaziz},
  \citenamefont {Ersen}, \citenamefont {Petit}, \citenamefont {Corbel},\ and\
  \citenamefont {Zafeiratos}}]{Derafa2018RSCA}%
  \BibitemOpen
  \bibfield  {author} {\bibinfo {author} {\bibfnamefont {W.}~\bibnamefont
  {Derafa}}, \bibinfo {author} {\bibfnamefont {F.}~\bibnamefont {Paloukis}},
  \bibinfo {author} {\bibfnamefont {B.}~\bibnamefont {Mewafy}}, \bibinfo
  {author} {\bibfnamefont {W.}~\bibnamefont {Baaziz}}, \bibinfo {author}
  {\bibfnamefont {O.}~\bibnamefont {Ersen}}, \bibinfo {author} {\bibfnamefont
  {C.}~\bibnamefont {Petit}}, \bibinfo {author} {\bibfnamefont
  {G.}~\bibnamefont {Corbel}},\ and\ \bibinfo {author} {\bibfnamefont
  {S.}~\bibnamefont {Zafeiratos}},\ }\bibfield  {title} {\bibinfo {title}
  {Synthesis and characterization of nickel-doped ceria nanoparticles with
  improved surface reducibility},\ }\href {https://doi.org/10.1039/C8RA07995A}
  {\bibfield  {journal} {\bibinfo  {journal} {RSC Adv.}\ }\textbf {\bibinfo
  {volume} {8}},\ \bibinfo {pages} {40712} (\bibinfo {year}
  {2018})}\BibitemShut {NoStop}%
\bibitem [{\citenamefont {Fuda}\ \emph {et~al.}(1984)\citenamefont {Fuda},
  \citenamefont {Kishio}, \citenamefont {Yamauchi}, \citenamefont {Fueki},\
  and\ \citenamefont {Onoda}}]{Fuda1984JPCS}%
  \BibitemOpen
  \bibfield  {author} {\bibinfo {author} {\bibfnamefont {K.}~\bibnamefont
  {Fuda}}, \bibinfo {author} {\bibfnamefont {K.}~\bibnamefont {Kishio}},
  \bibinfo {author} {\bibfnamefont {S.}~\bibnamefont {Yamauchi}}, \bibinfo
  {author} {\bibfnamefont {K.}~\bibnamefont {Fueki}},\ and\ \bibinfo {author}
  {\bibfnamefont {Y.}~\bibnamefont {Onoda}},\ }\bibfield  {title} {\bibinfo
  {title} {{$^{17}$O NMR study of Y$_2$O$_3$-doped CeO$_2$}},\ }\href
  {https://doi.org/10.1016/0022-3697(84)90024-6} {\bibfield  {journal}
  {\bibinfo  {journal} {J. Phys. Chem. Solids}\ }\textbf {\bibinfo {volume}
  {45}},\ \bibinfo {pages} {1253} (\bibinfo {year} {1984})}\BibitemShut
  {NoStop}%
\bibitem [{\citenamefont {Wang}\ \emph {et~al.}(1981)\citenamefont {Wang},
  \citenamefont {Park}, \citenamefont {Griffith},\ and\ \citenamefont
  {Nowick}}]{Wang1981SSI}%
  \BibitemOpen
  \bibfield  {author} {\bibinfo {author} {\bibfnamefont {D.~Y.}\ \bibnamefont
  {Wang}}, \bibinfo {author} {\bibfnamefont {D.}~\bibnamefont {Park}}, \bibinfo
  {author} {\bibfnamefont {J.}~\bibnamefont {Griffith}},\ and\ \bibinfo
  {author} {\bibfnamefont {A.}~\bibnamefont {Nowick}},\ }\bibfield  {title}
  {\bibinfo {title} {Oxygen-ion conductivity and defect interactions in
  yttria-doped ceria},\ }\href {https://doi.org/10.1016/0167-2738(81)90005-9}
  {\bibfield  {journal} {\bibinfo  {journal} {Solid State Ion.}\ }\textbf
  {\bibinfo {volume} {2}},\ \bibinfo {pages} {95} (\bibinfo {year}
  {1981})}\BibitemShut {NoStop}%
\bibitem [{\citenamefont {Lee}\ \emph {et~al.}(2014)\citenamefont {Lee},
  \citenamefont {Chen}, \citenamefont {Chen}, \citenamefont {Dong},
  \citenamefont {Lin}, \citenamefont {Chen},\ and\ \citenamefont
  {Gloter}}]{Lee2014JPCC}%
  \BibitemOpen
  \bibfield  {author} {\bibinfo {author} {\bibfnamefont {W.}~\bibnamefont
  {Lee}}, \bibinfo {author} {\bibfnamefont {S.-Y.}\ \bibnamefont {Chen}},
  \bibinfo {author} {\bibfnamefont {Y.-S.}\ \bibnamefont {Chen}}, \bibinfo
  {author} {\bibfnamefont {C.-L.}\ \bibnamefont {Dong}}, \bibinfo {author}
  {\bibfnamefont {H.-J.}\ \bibnamefont {Lin}}, \bibinfo {author} {\bibfnamefont
  {C.-T.}\ \bibnamefont {Chen}},\ and\ \bibinfo {author} {\bibfnamefont
  {A.}~\bibnamefont {Gloter}},\ }\bibfield  {title} {\bibinfo {title} {{Defect
  Structure Guided Room Temperature Ferromagnetism of Y-Doped CeO$_2$
  Nanoparticles}},\ }\href {https://doi.org/10.1021/jp507694d} {\bibfield
  {journal} {\bibinfo  {journal} {J. Phys. Chem. C}\ }\textbf {\bibinfo
  {volume} {118}},\ \bibinfo {pages} {26359} (\bibinfo {year}
  {2014})}\BibitemShut {NoStop}%
\bibitem [{\citenamefont {Skorodumova}\ \emph {et~al.}(2002)\citenamefont
  {Skorodumova}, \citenamefont {Simak}, \citenamefont {Lundqvist},
  \citenamefont {Abrikosov},\ and\ \citenamefont
  {Johansson}}]{Skorodumova2002PRL}%
  \BibitemOpen
  \bibfield  {author} {\bibinfo {author} {\bibfnamefont {N.~V.}\ \bibnamefont
  {Skorodumova}}, \bibinfo {author} {\bibfnamefont {S.~I.}\ \bibnamefont
  {Simak}}, \bibinfo {author} {\bibfnamefont {B.~I.}\ \bibnamefont
  {Lundqvist}}, \bibinfo {author} {\bibfnamefont {I.~A.}\ \bibnamefont
  {Abrikosov}},\ and\ \bibinfo {author} {\bibfnamefont {B.}~\bibnamefont
  {Johansson}},\ }\bibfield  {title} {\bibinfo {title} {{Quantum Origin of the
  Oxygen Storage Capability of Ceria}},\ }\href
  {https://doi.org/10.1103/PhysRevLett.89.166601} {\bibfield  {journal}
  {\bibinfo  {journal} {Phys. Rev. Lett.}\ }\textbf {\bibinfo {volume} {89}},\
  \bibinfo {pages} {166601} (\bibinfo {year} {2002})}\BibitemShut {NoStop}%
\bibitem [{\citenamefont {Keating}\ \emph {et~al.}(2012)\citenamefont
  {Keating}, \citenamefont {Scanlon}, \citenamefont {Morgan}, \citenamefont
  {Galea},\ and\ \citenamefont {Watson}}]{Keating2012JPCC}%
  \BibitemOpen
  \bibfield  {author} {\bibinfo {author} {\bibfnamefont {P.~R.~L.}\
  \bibnamefont {Keating}}, \bibinfo {author} {\bibfnamefont {D.~O.}\
  \bibnamefont {Scanlon}}, \bibinfo {author} {\bibfnamefont {B.~J.}\
  \bibnamefont {Morgan}}, \bibinfo {author} {\bibfnamefont {N.~M.}\
  \bibnamefont {Galea}},\ and\ \bibinfo {author} {\bibfnamefont {G.~W.}\
  \bibnamefont {Watson}},\ }\bibfield  {title} {\bibinfo {title} {{Analysis of
  Intrinsic Defects in CeO$_2$ Using a Koopmans-Like GGA$+$$U$ Approach}},\
  }\href {https://doi.org/10.1021/jp2080034} {\bibfield  {journal} {\bibinfo
  {journal} {J. Phys. Chem. C}\ }\textbf {\bibinfo {volume} {116}},\ \bibinfo
  {pages} {2443} (\bibinfo {year} {2012})}\BibitemShut {NoStop}%
\bibitem [{\citenamefont {Zacherle}\ \emph {et~al.}(2013)\citenamefont
  {Zacherle}, \citenamefont {Schriever}, \citenamefont {De~Souza},\ and\
  \citenamefont {Martin}}]{Zacherle2013PRB}%
  \BibitemOpen
  \bibfield  {author} {\bibinfo {author} {\bibfnamefont {T.}~\bibnamefont
  {Zacherle}}, \bibinfo {author} {\bibfnamefont {A.}~\bibnamefont {Schriever}},
  \bibinfo {author} {\bibfnamefont {R.~A.}\ \bibnamefont {De~Souza}},\ and\
  \bibinfo {author} {\bibfnamefont {M.}~\bibnamefont {Martin}},\ }\bibfield
  {title} {\bibinfo {title} {Ab initio analysis of the defect structure of
  ceria},\ }\href {https://doi.org/10.1103/PhysRevB.87.134104} {\bibfield
  {journal} {\bibinfo  {journal} {Phys. Rev. B}\ }\textbf {\bibinfo {volume}
  {87}},\ \bibinfo {pages} {134104} (\bibinfo {year} {2013})}\BibitemShut
  {NoStop}%
\bibitem [{\citenamefont {Huang}\ \emph {et~al.}(2014)\citenamefont {Huang},
  \citenamefont {Gillen},\ and\ \citenamefont {Robertson}}]{Huang2014JPCC}%
  \BibitemOpen
  \bibfield  {author} {\bibinfo {author} {\bibfnamefont {B.}~\bibnamefont
  {Huang}}, \bibinfo {author} {\bibfnamefont {R.}~\bibnamefont {Gillen}},\ and\
  \bibinfo {author} {\bibfnamefont {J.}~\bibnamefont {Robertson}},\ }\bibfield
  {title} {\bibinfo {title} {{Study of CeO$_2$ and Its Native Defects by
  Density Functional Theory with Repulsive Potential}},\ }\href
  {https://doi.org/10.1021/jp506625h} {\bibfield  {journal} {\bibinfo
  {journal} {J. Phys. Chem. C}\ }\textbf {\bibinfo {volume} {118}},\ \bibinfo
  {pages} {24248} (\bibinfo {year} {2014})}\BibitemShut {NoStop}%
\bibitem [{\citenamefont {Plata}\ \emph {et~al.}(2013)\citenamefont {Plata},
  \citenamefont {Márquez},\ and\ \citenamefont {Sanz}}]{Plata2013JPCC}%
  \BibitemOpen
  \bibfield  {author} {\bibinfo {author} {\bibfnamefont {J.~J.}\ \bibnamefont
  {Plata}}, \bibinfo {author} {\bibfnamefont {A.~M.}\ \bibnamefont
  {Márquez}},\ and\ \bibinfo {author} {\bibfnamefont {J.~F.}\ \bibnamefont
  {Sanz}},\ }\bibfield  {title} {\bibinfo {title} {{Electron Mobility via
  Polaron Hopping in Bulk Ceria: A First-Principles Study}},\ }\href
  {https://doi.org/10.1021/jp402594x} {\bibfield  {journal} {\bibinfo
  {journal} {J. Phys. Chem. C}\ }\textbf {\bibinfo {volume} {117}},\ \bibinfo
  {pages} {14502} (\bibinfo {year} {2013})}\BibitemShut {NoStop}%
\bibitem [{\citenamefont {Sun}\ \emph {et~al.}(2017)\citenamefont {Sun},
  \citenamefont {Huang}, \citenamefont {Wang},\ and\ \citenamefont
  {Janotti}}]{Sun2017PRB}%
  \BibitemOpen
  \bibfield  {author} {\bibinfo {author} {\bibfnamefont {L.}~\bibnamefont
  {Sun}}, \bibinfo {author} {\bibfnamefont {X.}~\bibnamefont {Huang}}, \bibinfo
  {author} {\bibfnamefont {L.}~\bibnamefont {Wang}},\ and\ \bibinfo {author}
  {\bibfnamefont {A.}~\bibnamefont {Janotti}},\ }\bibfield  {title} {\bibinfo
  {title} {Disentangling the role of small polarons and oxygen vacancies in
  {$\mathrm{Ce}{\mathrm{O}}_{2}$}},\ }\href
  {https://doi.org/10.1103/PhysRevB.95.245101} {\bibfield  {journal} {\bibinfo
  {journal} {Phys. Rev. B}\ }\textbf {\bibinfo {volume} {95}},\ \bibinfo
  {pages} {245101} (\bibinfo {year} {2017})}\BibitemShut {NoStop}%
\bibitem [{\citenamefont {Sun}\ and\ \citenamefont
  {Yildiz}(2019)}]{Sun2019JPCC}%
  \BibitemOpen
  \bibfield  {author} {\bibinfo {author} {\bibfnamefont {L.}~\bibnamefont
  {Sun}}\ and\ \bibinfo {author} {\bibfnamefont {B.}~\bibnamefont {Yildiz}},\
  }\bibfield  {title} {\bibinfo {title} {{Solubility Limit of Cu and Factors
  Governing the Reactivity of Cu–CeO$_2$ Assessed from First-Principles
  Defect Chemistry and Thermodynamics}},\ }\href
  {https://doi.org/10.1021/acs.jpcc.8b08222} {\bibfield  {journal} {\bibinfo
  {journal} {J. Phys. Chem. C}\ }\textbf {\bibinfo {volume} {123}},\ \bibinfo
  {pages} {399} (\bibinfo {year} {2019})}\BibitemShut {NoStop}%
\bibitem [{\citenamefont {Sun}\ \emph {et~al.}(2019)\citenamefont {Sun},
  \citenamefont {Hao}, \citenamefont {Meng}, \citenamefont {Wang},
  \citenamefont {Liu},\ and\ \citenamefont {Zhou}}]{Sun2019AEM}%
  \BibitemOpen
  \bibfield  {author} {\bibinfo {author} {\bibfnamefont {L.}~\bibnamefont
  {Sun}}, \bibinfo {author} {\bibfnamefont {X.}~\bibnamefont {Hao}}, \bibinfo
  {author} {\bibfnamefont {Q.}~\bibnamefont {Meng}}, \bibinfo {author}
  {\bibfnamefont {L.}~\bibnamefont {Wang}}, \bibinfo {author} {\bibfnamefont
  {F.}~\bibnamefont {Liu}},\ and\ \bibinfo {author} {\bibfnamefont
  {M.}~\bibnamefont {Zhou}},\ }\bibfield  {title} {\bibinfo {title} {{Polaronic
  Resistive Switching in Ceria-Based Memory Devices}},\ }\href
  {https://doi.org/10.1002/aelm.201900271} {\bibfield  {journal} {\bibinfo
  {journal} {Adv. Electron. Mater.}\ }\textbf {\bibinfo {volume} {5}},\
  \bibinfo {pages} {1900271} (\bibinfo {year} {2019})}\BibitemShut {NoStop}%
\bibitem [{\citenamefont {Zhang}\ \emph {et~al.}(2023)\citenamefont {Zhang},
  \citenamefont {Zhu}, \citenamefont {Hou}, \citenamefont {Guan}, \citenamefont
  {Lu}, \citenamefont {Keal}, \citenamefont {Buckeridge}, \citenamefont
  {Catlow},\ and\ \citenamefont {Sokol}}]{Zhang2023CM}%
  \BibitemOpen
  \bibfield  {author} {\bibinfo {author} {\bibfnamefont {X.}~\bibnamefont
  {Zhang}}, \bibinfo {author} {\bibfnamefont {L.}~\bibnamefont {Zhu}}, \bibinfo
  {author} {\bibfnamefont {Q.}~\bibnamefont {Hou}}, \bibinfo {author}
  {\bibfnamefont {J.}~\bibnamefont {Guan}}, \bibinfo {author} {\bibfnamefont
  {Y.}~\bibnamefont {Lu}}, \bibinfo {author} {\bibfnamefont {T.~W.}\
  \bibnamefont {Keal}}, \bibinfo {author} {\bibfnamefont {J.}~\bibnamefont
  {Buckeridge}}, \bibinfo {author} {\bibfnamefont {C.~R.~A.}\ \bibnamefont
  {Catlow}},\ and\ \bibinfo {author} {\bibfnamefont {A.~A.}\ \bibnamefont
  {Sokol}},\ }\bibfield  {title} {\bibinfo {title} {{Toward a Consistent
  Prediction of Defect Chemistry in CeO$_2$}},\ }\href
  {https://doi.org/10.1021/acs.chemmater.2c03019} {\bibfield  {journal}
  {\bibinfo  {journal} {Chem. Mater.}\ }\textbf {\bibinfo {volume} {35}},\
  \bibinfo {pages} {207} (\bibinfo {year} {2023})}\BibitemShut {NoStop}%
\bibitem [{\citenamefont {Sohlberg}\ \emph {et~al.}(2001)\citenamefont
  {Sohlberg}, \citenamefont {Pantelides},\ and\ \citenamefont
  {Pennycook}}]{Sohlberg2001JACS}%
  \BibitemOpen
  \bibfield  {author} {\bibinfo {author} {\bibfnamefont {K.}~\bibnamefont
  {Sohlberg}}, \bibinfo {author} {\bibfnamefont {S.~T.}\ \bibnamefont
  {Pantelides}},\ and\ \bibinfo {author} {\bibfnamefont {S.~J.}\ \bibnamefont
  {Pennycook}},\ }\bibfield  {title} {\bibinfo {title} {{Interactions of
  Hydrogen with CeO$_2$}},\ }\href {https://doi.org/10.1021/ja004008k}
  {\bibfield  {journal} {\bibinfo  {journal} {J. Am. Chem. Soc.}\ }\textbf
  {\bibinfo {volume} {123}},\ \bibinfo {pages} {6609} (\bibinfo {year}
  {2001})}\BibitemShut {NoStop}%
\bibitem [{\citenamefont {Stimac}\ and\ \citenamefont
  {Goldman}(2025)}]{Stimac2025Omega}%
  \BibitemOpen
  \bibfield  {author} {\bibinfo {author} {\bibfnamefont {J.~C.}\ \bibnamefont
  {Stimac}}\ and\ \bibinfo {author} {\bibfnamefont {N.}~\bibnamefont
  {Goldman}},\ }\bibfield  {title} {\bibinfo {title} {{Quantum Calculations of
  Hydrogen Absorption and Diffusivity in Bulk CeO$_2$}},\ }\href
  {https://doi.org/10.1021/acsomega.4c11470} {\bibfield  {journal} {\bibinfo
  {journal} {ACS Omega}\ }\textbf {\bibinfo {volume} {10}},\ \bibinfo {pages}
  {12385} (\bibinfo {year} {2025})}\BibitemShut {NoStop}%
\bibitem [{\citenamefont {Wang}\ \emph {et~al.}(2010)\citenamefont {Wang},
  \citenamefont {Shen}, \citenamefont {Wang},\ and\ \citenamefont
  {Fabris}}]{Wang2010JPCC}%
  \BibitemOpen
  \bibfield  {author} {\bibinfo {author} {\bibfnamefont {X.}~\bibnamefont
  {Wang}}, \bibinfo {author} {\bibfnamefont {M.}~\bibnamefont {Shen}}, \bibinfo
  {author} {\bibfnamefont {J.}~\bibnamefont {Wang}},\ and\ \bibinfo {author}
  {\bibfnamefont {S.}~\bibnamefont {Fabris}},\ }\bibfield  {title} {\bibinfo
  {title} {{Enhanced Oxygen Buffering by Substitutional and Interstitial Ni
  Point Defects in Ceria: A First-Principles DFT$+$$U$ Study}},\ }\href
  {https://doi.org/10.1021/jp101100f} {\bibfield  {journal} {\bibinfo
  {journal} {J. Phys. Chem. C}\ }\textbf {\bibinfo {volume} {114}},\ \bibinfo
  {pages} {10221} (\bibinfo {year} {2010})}\BibitemShut {NoStop}%
\bibitem [{\citenamefont {Anisimov}\ \emph {et~al.}(1991)\citenamefont
  {Anisimov}, \citenamefont {Zaanen},\ and\ \citenamefont
  {Andersen}}]{anisimov1991}%
  \BibitemOpen
  \bibfield  {author} {\bibinfo {author} {\bibfnamefont {V.~I.}\ \bibnamefont
  {Anisimov}}, \bibinfo {author} {\bibfnamefont {J.}~\bibnamefont {Zaanen}},\
  and\ \bibinfo {author} {\bibfnamefont {O.~K.}\ \bibnamefont {Andersen}},\
  }\bibfield  {title} {\bibinfo {title} {{Band theory and Mott insulators:
  Hubbard U instead of Stoner I}},\ }\href
  {https://doi.org/10.1103/PhysRevB.44.943} {\bibfield  {journal} {\bibinfo
  {journal} {Phys. Rev. B}\ }\textbf {\bibinfo {volume} {44}},\ \bibinfo
  {pages} {943} (\bibinfo {year} {1991})}\BibitemShut {NoStop}%
\bibitem [{\citenamefont {Liechtenstein}\ \emph {et~al.}(1995)\citenamefont
  {Liechtenstein}, \citenamefont {Anisimov},\ and\ \citenamefont
  {Zaanen}}]{liechtenstein1995}%
  \BibitemOpen
  \bibfield  {author} {\bibinfo {author} {\bibfnamefont {A.~I.}\ \bibnamefont
  {Liechtenstein}}, \bibinfo {author} {\bibfnamefont {V.~I.}\ \bibnamefont
  {Anisimov}},\ and\ \bibinfo {author} {\bibfnamefont {J.}~\bibnamefont
  {Zaanen}},\ }\bibfield  {title} {\bibinfo {title} {{Density-functional theory
  and strong interactions: Orbital ordering in Mott-Hubbard insulators}},\
  }\href {https://doi.org/10.1103/PhysRevB.52.R5467} {\bibfield  {journal}
  {\bibinfo  {journal} {Phys. Rev. B}\ }\textbf {\bibinfo {volume} {52}},\
  \bibinfo {pages} {R5467} (\bibinfo {year} {1995})}\BibitemShut {NoStop}%
\bibitem [{\citenamefont {Heyd}\ \emph {et~al.}(2003)\citenamefont {Heyd},
  \citenamefont {Scuseria},\ and\ \citenamefont {Ernzerhof}}]{heyd:8207}%
  \BibitemOpen
  \bibfield  {author} {\bibinfo {author} {\bibfnamefont {J.}~\bibnamefont
  {Heyd}}, \bibinfo {author} {\bibfnamefont {G.~E.}\ \bibnamefont {Scuseria}},\
  and\ \bibinfo {author} {\bibfnamefont {M.}~\bibnamefont {Ernzerhof}},\
  }\bibfield  {title} {\bibinfo {title} {Hybrid functionals based on a screened
  {C}oulomb potential},\ }\href {https://doi.org/10.1063/1.1564060} {\bibfield
  {journal} {\bibinfo  {journal} {J. Chem. Phys.}\ }\textbf {\bibinfo {volume}
  {118}},\ \bibinfo {pages} {8207} (\bibinfo {year} {2003})}\BibitemShut
  {NoStop}%
\bibitem [{\citenamefont {Paier}\ \emph {et~al.}(2006)\citenamefont {Paier},
  \citenamefont {Marsman}, \citenamefont {Hummer}, \citenamefont {Kresse},
  \citenamefont {Gerber},\ and\ \citenamefont {\'{A}ngy\'{a}n}}]{paier:154709}%
  \BibitemOpen
  \bibfield  {author} {\bibinfo {author} {\bibfnamefont {J.}~\bibnamefont
  {Paier}}, \bibinfo {author} {\bibfnamefont {M.}~\bibnamefont {Marsman}},
  \bibinfo {author} {\bibfnamefont {K.}~\bibnamefont {Hummer}}, \bibinfo
  {author} {\bibfnamefont {G.}~\bibnamefont {Kresse}}, \bibinfo {author}
  {\bibfnamefont {I.~C.}\ \bibnamefont {Gerber}},\ and\ \bibinfo {author}
  {\bibfnamefont {J.~G.}\ \bibnamefont {\'{A}ngy\'{a}n}},\ }\bibfield  {title}
  {\bibinfo {title} {Screened hybrid density functionals applied to solids},\
  }\href {https://doi.org/10.1063/1.2187006} {\bibfield  {journal} {\bibinfo
  {journal} {J. Chem. Phys.}\ }\textbf {\bibinfo {volume} {124}},\ \bibinfo
  {eid} {154709} (\bibinfo {year} {2006})}\BibitemShut {NoStop}%
\bibitem [{\citenamefont {Bl\"ochl}(1994)}]{PAW1}%
  \BibitemOpen
  \bibfield  {author} {\bibinfo {author} {\bibfnamefont {P.~E.}\ \bibnamefont
  {Bl\"ochl}},\ }\bibfield  {title} {\bibinfo {title} {Projector augmented-wave
  method},\ }\href {https://doi.org/10.1103/PhysRevB.50.17953} {\bibfield
  {journal} {\bibinfo  {journal} {Phys. Rev. B}\ }\textbf {\bibinfo {volume}
  {50}},\ \bibinfo {pages} {17953} (\bibinfo {year} {1994})}\BibitemShut
  {NoStop}%
\bibitem [{\citenamefont {Kresse}\ and\ \citenamefont {Hafner}(1993)}]{VASP1}%
  \BibitemOpen
  \bibfield  {author} {\bibinfo {author} {\bibfnamefont {G.}~\bibnamefont
  {Kresse}}\ and\ \bibinfo {author} {\bibfnamefont {J.}~\bibnamefont
  {Hafner}},\ }\bibfield  {title} {\bibinfo {title} {Ab initio molecular
  dynamics for liquid metals},\ }\href
  {https://doi.org/10.1103/PhysRevB.47.558} {\bibfield  {journal} {\bibinfo
  {journal} {Phys. Rev. B}\ }\textbf {\bibinfo {volume} {47}},\ \bibinfo
  {pages} {558} (\bibinfo {year} {1993})}\BibitemShut {NoStop}%
\bibitem [{\citenamefont {Kresse}\ and\ \citenamefont
  {Furthm\"uller}(1996{\natexlab{a}})}]{VASP2}%
  \BibitemOpen
  \bibfield  {author} {\bibinfo {author} {\bibfnamefont {G.}~\bibnamefont
  {Kresse}}\ and\ \bibinfo {author} {\bibfnamefont {J.}~\bibnamefont
  {Furthm\"uller}},\ }\bibfield  {title} {\bibinfo {title} {Efficient iterative
  schemes for ab initio total-energy calculations using a plane-wave basis
  set},\ }\href {https://doi.org/10.1103/PhysRevB.54.11169} {\bibfield
  {journal} {\bibinfo  {journal} {Phys. Rev. B}\ }\textbf {\bibinfo {volume}
  {54}},\ \bibinfo {pages} {11169} (\bibinfo {year}
  {1996}{\natexlab{a}})}\BibitemShut {NoStop}%
\bibitem [{\citenamefont {Kresse}\ and\ \citenamefont
  {Furthm\"uller}(1996{\natexlab{b}})}]{VASP3}%
  \BibitemOpen
  \bibfield  {author} {\bibinfo {author} {\bibfnamefont {G.}~\bibnamefont
  {Kresse}}\ and\ \bibinfo {author} {\bibfnamefont {J.}~\bibnamefont
  {Furthm\"uller}},\ }\bibfield  {title} {\bibinfo {title} {Efficiency of
  ab-initio total energy calculations for metals and semiconductors using a
  plane-wave basis set},\ }\href {https://doi.org/DOI:
  10.1016/0927-0256(96)00008-0} {\bibfield  {journal} {\bibinfo  {journal}
  {Comput. Mat. Sci.}\ }\textbf {\bibinfo {volume} {6}},\ \bibinfo {pages} {15}
  (\bibinfo {year} {1996}{\natexlab{b}})}\BibitemShut {NoStop}%
\bibitem [{\citenamefont {{Van de Walle}}\ and\ \citenamefont
  {Neugebauer}(2004)}]{walle:3851}%
  \BibitemOpen
  \bibfield  {author} {\bibinfo {author} {\bibfnamefont {C.~G.}\ \bibnamefont
  {{Van de Walle}}}\ and\ \bibinfo {author} {\bibfnamefont {J.}~\bibnamefont
  {Neugebauer}},\ }\bibfield  {title} {\bibinfo {title} {First-principles
  calculations for defects and impurities: Applications to {III}-nitrides},\
  }\href {https://doi.org/10.1063/1.1682673} {\bibfield  {journal} {\bibinfo
  {journal} {J. Appl. Phys.}\ }\textbf {\bibinfo {volume} {95}},\ \bibinfo
  {pages} {3851} (\bibinfo {year} {2004})}\BibitemShut {NoStop}%
\bibitem [{\citenamefont {Freysoldt}\ \emph {et~al.}(2014)\citenamefont
  {Freysoldt}, \citenamefont {Grabowski}, \citenamefont {Hickel}, \citenamefont
  {Neugebauer}, \citenamefont {Kresse}, \citenamefont {Janotti},\ and\
  \citenamefont {{Van de Walle}}}]{Freysoldt2014RMP}%
  \BibitemOpen
  \bibfield  {author} {\bibinfo {author} {\bibfnamefont {C.}~\bibnamefont
  {Freysoldt}}, \bibinfo {author} {\bibfnamefont {B.}~\bibnamefont
  {Grabowski}}, \bibinfo {author} {\bibfnamefont {T.}~\bibnamefont {Hickel}},
  \bibinfo {author} {\bibfnamefont {J.}~\bibnamefont {Neugebauer}}, \bibinfo
  {author} {\bibfnamefont {G.}~\bibnamefont {Kresse}}, \bibinfo {author}
  {\bibfnamefont {A.}~\bibnamefont {Janotti}},\ and\ \bibinfo {author}
  {\bibfnamefont {C.~G.}\ \bibnamefont {{Van de Walle}}},\ }\bibfield  {title}
  {\bibinfo {title} {First-principles calculations for point defects in
  solids},\ }\href {https://doi.org/10.1103/RevModPhys.86.253} {\bibfield
  {journal} {\bibinfo  {journal} {Rev. Mod. Phys.}\ }\textbf {\bibinfo {volume}
  {86}},\ \bibinfo {pages} {253} (\bibinfo {year} {2014})}\BibitemShut
  {NoStop}%
\bibitem [{\citenamefont {Freysoldt}\ \emph {et~al.}(2009)\citenamefont
  {Freysoldt}, \citenamefont {Neugebauer},\ and\ \citenamefont {{Van de
  Walle}}}]{Freysoldt}%
  \BibitemOpen
  \bibfield  {author} {\bibinfo {author} {\bibfnamefont {C.}~\bibnamefont
  {Freysoldt}}, \bibinfo {author} {\bibfnamefont {J.}~\bibnamefont
  {Neugebauer}},\ and\ \bibinfo {author} {\bibfnamefont {C.~G.}\ \bibnamefont
  {{Van de Walle}}},\ }\bibfield  {title} {\bibinfo {title} {Fully ab initio
  finite-size corrections for charged-defect supercell calculations},\ }\href
  {https://doi.org/10.1103/PhysRevLett.102.016402} {\bibfield  {journal}
  {\bibinfo  {journal} {Phys. Rev. Lett.}\ }\textbf {\bibinfo {volume} {102}},\
  \bibinfo {pages} {016402} (\bibinfo {year} {2009})}\BibitemShut {NoStop}%
\bibitem [{\citenamefont {{Chase, Jr.}}(1998)}]{chase}%
  \BibitemOpen
  \bibfield  {author} {\bibinfo {author} {\bibfnamefont {M.~W.}\ \bibnamefont
  {{Chase, Jr.}}},\ }\href@noop {} {\emph {\bibinfo {title} {NIST-JANAF
  Themochemical Tables, Fourth Edition}}}\ (\bibinfo  {publisher} {J. Phys.
  Chem. Ref. Data, Monograph 9},\ \bibinfo {year} {1998})\ pp.\ \bibinfo
  {pages} {1--1951}\BibitemShut {NoStop}%
\bibitem [{\citenamefont {Rosso}\ \emph {et~al.}(2003)\citenamefont {Rosso},
  \citenamefont {Smith},\ and\ \citenamefont {Dupuis}}]{Rosso2003}%
  \BibitemOpen
  \bibfield  {author} {\bibinfo {author} {\bibfnamefont {K.~M.}\ \bibnamefont
  {Rosso}}, \bibinfo {author} {\bibfnamefont {D.~M.~A.}\ \bibnamefont
  {Smith}},\ and\ \bibinfo {author} {\bibfnamefont {M.}~\bibnamefont
  {Dupuis}},\ }\bibfield  {title} {\bibinfo {title} {An ab initio model of
  electron transport in hematite ($\alpha$-{Fe$_2$O$_3$}) basal planes},\
  }\href {https://doi.org/10.1063/1.1558534} {\bibfield  {journal} {\bibinfo
  {journal} {J. Chem. Phys.}\ }\textbf {\bibinfo {volume} {118}},\ \bibinfo
  {pages} {6455} (\bibinfo {year} {2003})}\BibitemShut {NoStop}%
\bibitem [{\citenamefont {Henkelman}\ \emph {et~al.}(2000)\citenamefont
  {Henkelman}, \citenamefont {Uberuaga},\ and\ \citenamefont
  {J\'{o}nsson}}]{ci-neb}%
  \BibitemOpen
  \bibfield  {author} {\bibinfo {author} {\bibfnamefont {G.}~\bibnamefont
  {Henkelman}}, \bibinfo {author} {\bibfnamefont {B.~P.}\ \bibnamefont
  {Uberuaga}},\ and\ \bibinfo {author} {\bibfnamefont {H.}~\bibnamefont
  {J\'{o}nsson}},\ }\bibfield  {title} {\bibinfo {title} {A climbing image
  nudged elastic band method for finding saddle points and minimum energy
  paths},\ }\href {https://doi.org/10.1063/1.1329672} {\bibfield  {journal}
  {\bibinfo  {journal} {J. Chem. Phys.}\ }\textbf {\bibinfo {volume} {113}},\
  \bibinfo {pages} {9901} (\bibinfo {year} {2000})}\BibitemShut {NoStop}%
\bibitem [{\citenamefont {Perdew}\ \emph {et~al.}(1996)\citenamefont {Perdew},
  \citenamefont {Burke},\ and\ \citenamefont {Ernzerhof}}]{GGA}%
  \BibitemOpen
  \bibfield  {author} {\bibinfo {author} {\bibfnamefont {J.~P.}\ \bibnamefont
  {Perdew}}, \bibinfo {author} {\bibfnamefont {K.}~\bibnamefont {Burke}},\ and\
  \bibinfo {author} {\bibfnamefont {M.}~\bibnamefont {Ernzerhof}},\ }\bibfield
  {title} {\bibinfo {title} {Generalized gradient approximation made simple},\
  }\href {https://doi.org/10.1103/PhysRevLett.77.3865} {\bibfield  {journal}
  {\bibinfo  {journal} {Phys. Rev. Lett.}\ }\textbf {\bibinfo {volume} {77}},\
  \bibinfo {pages} {3865} (\bibinfo {year} {1996})}\BibitemShut {NoStop}%
\bibitem [{\citenamefont {Castleton}\ \emph {et~al.}(2019)\citenamefont
  {Castleton}, \citenamefont {Lee},\ and\ \citenamefont
  {Kullgren}}]{Castleton2019JPCC}%
  \BibitemOpen
  \bibfield  {author} {\bibinfo {author} {\bibfnamefont {C.~W.~M.}\
  \bibnamefont {Castleton}}, \bibinfo {author} {\bibfnamefont {A.}~\bibnamefont
  {Lee}},\ and\ \bibinfo {author} {\bibfnamefont {J.}~\bibnamefont
  {Kullgren}},\ }\bibfield  {title} {\bibinfo {title} {{Benchmarking Density
  Functional Theory Functionals for Polarons in Oxides: Properties of
  CeO$_2$}},\ }\href {https://doi.org/10.1021/acs.jpcc.8b09134} {\bibfield
  {journal} {\bibinfo  {journal} {J. Phys. Chem. C}\ }\textbf {\bibinfo
  {volume} {123}},\ \bibinfo {pages} {5164} (\bibinfo {year}
  {2019})}\BibitemShut {NoStop}%
\bibitem [{\citenamefont {Hoang}(2021)}]{Hoang2021PRM}%
  \BibitemOpen
  \bibfield  {author} {\bibinfo {author} {\bibfnamefont {K.}~\bibnamefont
  {Hoang}},\ }\bibfield  {title} {\bibinfo {title} {Tuning the valence and
  concentration of europium and luminescence centers in {GaN} through co-doping
  and defect association},\ }\href
  {https://doi.org/10.1103/PhysRevMaterials.5.034601} {\bibfield  {journal}
  {\bibinfo  {journal} {Phys. Rev. Materials}\ }\textbf {\bibinfo {volume}
  {5}},\ \bibinfo {pages} {034601} (\bibinfo {year} {2021})}\BibitemShut
  {NoStop}%
\bibitem [{\citenamefont {Gupta}\ and\ \citenamefont
  {Singh}(1970)}]{Gupta1970JACeS}%
  \BibitemOpen
  \bibfield  {author} {\bibinfo {author} {\bibfnamefont {M.~L.}\ \bibnamefont
  {Gupta}}\ and\ \bibinfo {author} {\bibfnamefont {S.}~\bibnamefont {Singh}},\
  }\bibfield  {title} {\bibinfo {title} {{Thermal Expansion of CeO$_2$,
  Ho$_2$O$_3$, and Lu$_2$O3 from 100$^\circ$ to 300$^\circ$K by an X-Ray
  Method}},\ }\href {https://doi.org/10.1111/j.1151-2916.1970.tb12037.x}
  {\bibfield  {journal} {\bibinfo  {journal} {J. Am. Ceram. Soc.}\ }\textbf
  {\bibinfo {volume} {53}},\ \bibinfo {pages} {663} (\bibinfo {year}
  {1970})}\BibitemShut {NoStop}%
\bibitem [{\citenamefont {Gajdo\ifmmode~\check{s}\else \v{s}\fi{}}\ \emph
  {et~al.}(2006)\citenamefont {Gajdo\ifmmode~\check{s}\else \v{s}\fi{}},
  \citenamefont {Hummer}, \citenamefont {Kresse}, \citenamefont
  {Furthm\"uller},\ and\ \citenamefont {Bechstedt}}]{Gajdos2006PRB}%
  \BibitemOpen
  \bibfield  {author} {\bibinfo {author} {\bibfnamefont {M.}~\bibnamefont
  {Gajdo\ifmmode~\check{s}\else \v{s}\fi{}}}, \bibinfo {author} {\bibfnamefont
  {K.}~\bibnamefont {Hummer}}, \bibinfo {author} {\bibfnamefont
  {G.}~\bibnamefont {Kresse}}, \bibinfo {author} {\bibfnamefont
  {J.}~\bibnamefont {Furthm\"uller}},\ and\ \bibinfo {author} {\bibfnamefont
  {F.}~\bibnamefont {Bechstedt}},\ }\bibfield  {title} {\bibinfo {title}
  {Linear optical properties in the projector-augmented wave methodology},\
  }\href {https://doi.org/10.1103/PhysRevB.73.045112} {\bibfield  {journal}
  {\bibinfo  {journal} {Phys. Rev. B}\ }\textbf {\bibinfo {volume} {73}},\
  \bibinfo {pages} {045112} (\bibinfo {year} {2006})}\BibitemShut {NoStop}%
\bibitem [{\citenamefont {Lany}(2024)}]{Lany2024JACS}%
  \BibitemOpen
  \bibfield  {author} {\bibinfo {author} {\bibfnamefont {S.}~\bibnamefont
  {Lany}},\ }\bibfield  {title} {\bibinfo {title} {{Chemical Potential Analysis
  as an Alternative to the van't Hoff Method: Hypothetical Limits of Solar
  Thermochemical Hydrogen}},\ }\href {https://doi.org/10.1021/jacs.4c02688}
  {\bibfield  {journal} {\bibinfo  {journal} {J. Am. Chem. Soc.}\ }\textbf
  {\bibinfo {volume} {146}},\ \bibinfo {pages} {14114} (\bibinfo {year}
  {2024})}\BibitemShut {NoStop}%
\bibitem [{\citenamefont {Momma}\ and\ \citenamefont {Izumi}(2011)}]{VESTA}%
  \BibitemOpen
  \bibfield  {author} {\bibinfo {author} {\bibfnamefont {K.}~\bibnamefont
  {Momma}}\ and\ \bibinfo {author} {\bibfnamefont {F.}~\bibnamefont {Izumi}},\
  }\bibfield  {title} {\bibinfo {title} {{{\it VESTA3} for three-dimensional
  visualization of crystal, volumetric and morphology data}},\ }\href
  {https://doi.org/10.1107/S0021889811038970} {\bibfield  {journal} {\bibinfo
  {journal} {J. Appl. Cryst.}\ }\textbf {\bibinfo {volume} {44}},\ \bibinfo
  {pages} {1272} (\bibinfo {year} {2011})}\BibitemShut {NoStop}%
\bibitem [{\citenamefont {Mamontov}\ and\ \citenamefont
  {Egami}(2000)}]{Mamontov2000JPCS}%
  \BibitemOpen
  \bibfield  {author} {\bibinfo {author} {\bibfnamefont {E.}~\bibnamefont
  {Mamontov}}\ and\ \bibinfo {author} {\bibfnamefont {T.}~\bibnamefont
  {Egami}},\ }\bibfield  {title} {\bibinfo {title} {Structural defects in a
  nano-scale powder of {CeO$_2$} studied by pulsed neutron diffraction},\
  }\href {https://doi.org/10.1016/S0022-3697(00)00003-2} {\bibfield  {journal}
  {\bibinfo  {journal} {J. Phys. Chem. Solids}\ }\textbf {\bibinfo {volume}
  {61}},\ \bibinfo {pages} {1345} (\bibinfo {year} {2000})}\BibitemShut
  {NoStop}%
\bibitem [{\citenamefont {Luo}\ \emph {et~al.}(2021)\citenamefont {Luo},
  \citenamefont {Li}, \citenamefont {Fung}, \citenamefont {Sumpter},
  \citenamefont {Liu}, \citenamefont {Wu},\ and\ \citenamefont
  {Page}}]{Luo2021CM}%
  \BibitemOpen
  \bibfield  {author} {\bibinfo {author} {\bibfnamefont {S.}~\bibnamefont
  {Luo}}, \bibinfo {author} {\bibfnamefont {M.}~\bibnamefont {Li}}, \bibinfo
  {author} {\bibfnamefont {V.}~\bibnamefont {Fung}}, \bibinfo {author}
  {\bibfnamefont {B.~G.}\ \bibnamefont {Sumpter}}, \bibinfo {author}
  {\bibfnamefont {J.}~\bibnamefont {Liu}}, \bibinfo {author} {\bibfnamefont
  {Z.}~\bibnamefont {Wu}},\ and\ \bibinfo {author} {\bibfnamefont
  {K.}~\bibnamefont {Page}},\ }\bibfield  {title} {\bibinfo {title} {{New
  Insights into the Bulk and Surface Defect Structures of Ceria Nanocrystals
  from Neutron Scattering Study}},\ }\href
  {https://doi.org/10.1021/acs.chemmater.1c00156} {\bibfield  {journal}
  {\bibinfo  {journal} {Chem. Mater.}\ }\textbf {\bibinfo {volume} {33}},\
  \bibinfo {pages} {3959} (\bibinfo {year} {2021})}\BibitemShut {NoStop}%
\bibitem [{\citenamefont {Wardenga}\ and\ \citenamefont
  {Klein}(2016)}]{Wardenga2016ASS}%
  \BibitemOpen
  \bibfield  {author} {\bibinfo {author} {\bibfnamefont {H.~F.}\ \bibnamefont
  {Wardenga}}\ and\ \bibinfo {author} {\bibfnamefont {A.}~\bibnamefont
  {Klein}},\ }\bibfield  {title} {\bibinfo {title} {Surface potentials of
  (111), (110) and (100) oriented {CeO$_{2-x}$} thin films},\ }\href
  {https://doi.org/10.1016/j.apsusc.2016.03.091} {\bibfield  {journal}
  {\bibinfo  {journal} {Appl. Surf. Sci.}\ }\textbf {\bibinfo {volume} {377}},\
  \bibinfo {pages} {1} (\bibinfo {year} {2016})}\BibitemShut {NoStop}%
\bibitem [{\citenamefont {Crovetto}\ \emph {et~al.}(2016)\citenamefont
  {Crovetto}, \citenamefont {Yan}, \citenamefont {Iandolo}, \citenamefont
  {Zhou}, \citenamefont {Stride}, \citenamefont {Schou}, \citenamefont {Hao},\
  and\ \citenamefont {Hansen}}]{Crovetto2016APL}%
  \BibitemOpen
  \bibfield  {author} {\bibinfo {author} {\bibfnamefont {A.}~\bibnamefont
  {Crovetto}}, \bibinfo {author} {\bibfnamefont {C.}~\bibnamefont {Yan}},
  \bibinfo {author} {\bibfnamefont {B.}~\bibnamefont {Iandolo}}, \bibinfo
  {author} {\bibfnamefont {F.}~\bibnamefont {Zhou}}, \bibinfo {author}
  {\bibfnamefont {J.}~\bibnamefont {Stride}}, \bibinfo {author} {\bibfnamefont
  {J.}~\bibnamefont {Schou}}, \bibinfo {author} {\bibfnamefont
  {X.}~\bibnamefont {Hao}},\ and\ \bibinfo {author} {\bibfnamefont
  {O.}~\bibnamefont {Hansen}},\ }\bibfield  {title} {\bibinfo {title}
  {Lattice-matched {Cu$_2$ZnSnS$_4$/CeO$_2$} solar cell with open circuit
  voltage boost},\ }\href {https://doi.org/10.1063/1.4971779} {\bibfield
  {journal} {\bibinfo  {journal} {Appl. Phys. Lett.}\ }\textbf {\bibinfo
  {volume} {109}},\ \bibinfo {pages} {233904} (\bibinfo {year}
  {2016})}\BibitemShut {NoStop}%
\bibitem [{\citenamefont {Tuller}\ and\ \citenamefont
  {Nowick}(1979)}]{Tuller1979JES}%
  \BibitemOpen
  \bibfield  {author} {\bibinfo {author} {\bibfnamefont {H.~L.}\ \bibnamefont
  {Tuller}}\ and\ \bibinfo {author} {\bibfnamefont {A.~S.}\ \bibnamefont
  {Nowick}},\ }\bibfield  {title} {\bibinfo {title} {{Defect Structure and
  Electrical Properties of Nonstoichiometric CeO$_2$ Single Crystals}},\ }\href
  {https://doi.org/10.1149/1.2129007} {\bibfield  {journal} {\bibinfo
  {journal} {J. Electrochem. Soc.}\ }\textbf {\bibinfo {volume} {126}},\
  \bibinfo {pages} {209} (\bibinfo {year} {1979})}\BibitemShut {NoStop}%
\bibitem [{\citenamefont {Nelson}\ \emph {et~al.}(2014)\citenamefont {Nelson},
  \citenamefont {Rittman}, \citenamefont {White}, \citenamefont {Dunwoody},
  \citenamefont {Kato},\ and\ \citenamefont {McClellan}}]{Nelson2014JACeS}%
  \BibitemOpen
  \bibfield  {author} {\bibinfo {author} {\bibfnamefont {A.~T.}\ \bibnamefont
  {Nelson}}, \bibinfo {author} {\bibfnamefont {D.~R.}\ \bibnamefont {Rittman}},
  \bibinfo {author} {\bibfnamefont {J.~T.}\ \bibnamefont {White}}, \bibinfo
  {author} {\bibfnamefont {J.~T.}\ \bibnamefont {Dunwoody}}, \bibinfo {author}
  {\bibfnamefont {M.}~\bibnamefont {Kato}},\ and\ \bibinfo {author}
  {\bibfnamefont {K.~J.}\ \bibnamefont {McClellan}},\ }\bibfield  {title}
  {\bibinfo {title} {{An Evaluation of the Thermophysical Properties of
  Stoichiometric CeO$_2$ in Comparison to UO$_2$ and PuO$_2$}},\ }\href
  {https://doi.org/10.1111/jace.13170} {\bibfield  {journal} {\bibinfo
  {journal} {J. Am. Ceram. Soc.}\ }\textbf {\bibinfo {volume} {97}},\ \bibinfo
  {pages} {3652} (\bibinfo {year} {2014})}\BibitemShut {NoStop}%
\bibitem [{\citenamefont {Huang}\ \emph {et~al.}(1998)\citenamefont {Huang},
  \citenamefont {Feng},\ and\ \citenamefont {Goodenough}}]{Huang1998JACeS}%
  \BibitemOpen
  \bibfield  {author} {\bibinfo {author} {\bibfnamefont {K.}~\bibnamefont
  {Huang}}, \bibinfo {author} {\bibfnamefont {M.}~\bibnamefont {Feng}},\ and\
  \bibinfo {author} {\bibfnamefont {J.~B.}\ \bibnamefont {Goodenough}},\
  }\bibfield  {title} {\bibinfo {title} {{Synthesis and Electrical Properties
  of Dense Ce$_{0.9}$Gd$_{0.1}$O$_{1.95}$ Ceramics}},\ }\href
  {https://doi.org/10.1111/j.1151-2916.1998.tb02341.x} {\bibfield  {journal}
  {\bibinfo  {journal} {J. Am. Ceram. Soc.}\ }\textbf {\bibinfo {volume}
  {81}},\ \bibinfo {pages} {357} (\bibinfo {year} {1998})}\BibitemShut
  {NoStop}%
\bibitem [{\citenamefont {Steele}(2000)}]{Steele2000SSI}%
  \BibitemOpen
  \bibfield  {author} {\bibinfo {author} {\bibfnamefont {B.}~\bibnamefont
  {Steele}},\ }\bibfield  {title} {\bibinfo {title} {{Appraisal of
  Ce$_{1-y}$Gd$_y$O$_{2-y/2}$ electrolytes for IT-SOFC operation at
  500$^\circ$C}},\ }\href {https://doi.org/10.1016/S0167-2738(99)00319-7}
  {\bibfield  {journal} {\bibinfo  {journal} {Solid State Ion.}\ }\textbf
  {\bibinfo {volume} {129}},\ \bibinfo {pages} {95} (\bibinfo {year}
  {2000})}\BibitemShut {NoStop}%
\bibitem [{\citenamefont {Lai}\ and\ \citenamefont
  {Haile}(2005)}]{Lai2005JACeS}%
  \BibitemOpen
  \bibfield  {author} {\bibinfo {author} {\bibfnamefont {W.}~\bibnamefont
  {Lai}}\ and\ \bibinfo {author} {\bibfnamefont {S.~M.}\ \bibnamefont
  {Haile}},\ }\bibfield  {title} {\bibinfo {title} {{Impedance Spectroscopy as
  a Tool for Chemical and Electrochemical Analysis of Mixed Conductors: A Case
  Study of Ceria}},\ }\href {https://doi.org/10.1111/j.1551-2916.2005.00740.x}
  {\bibfield  {journal} {\bibinfo  {journal} {J. Am. Ceram. Soc.}\ }\textbf
  {\bibinfo {volume} {88}},\ \bibinfo {pages} {2979} (\bibinfo {year}
  {2005})}\BibitemShut {NoStop}%
\bibitem [{\citenamefont {Sakai}\ \emph {et~al.}(1999)\citenamefont {Sakai},
  \citenamefont {Yamaji}, \citenamefont {Horita}, \citenamefont {Yokokawa},
  \citenamefont {Hirata}, \citenamefont {Sameshima}, \citenamefont {Nigara},\
  and\ \citenamefont {Mizusaki}}]{Sakai1999SSI}%
  \BibitemOpen
  \bibfield  {author} {\bibinfo {author} {\bibfnamefont {N.}~\bibnamefont
  {Sakai}}, \bibinfo {author} {\bibfnamefont {K.}~\bibnamefont {Yamaji}},
  \bibinfo {author} {\bibfnamefont {T.}~\bibnamefont {Horita}}, \bibinfo
  {author} {\bibfnamefont {H.}~\bibnamefont {Yokokawa}}, \bibinfo {author}
  {\bibfnamefont {Y.}~\bibnamefont {Hirata}}, \bibinfo {author} {\bibfnamefont
  {S.}~\bibnamefont {Sameshima}}, \bibinfo {author} {\bibfnamefont
  {Y.}~\bibnamefont {Nigara}},\ and\ \bibinfo {author} {\bibfnamefont
  {J.}~\bibnamefont {Mizusaki}},\ }\bibfield  {title} {\bibinfo {title}
  {Determination of hydrogen solubility in oxide ceramics by using {SIMS}
  analyses},\ }\href {https://doi.org/10.1016/S0167-2738(99)00192-7} {\bibfield
   {journal} {\bibinfo  {journal} {Solid State Ion.}\ }\textbf {\bibinfo
  {volume} {125}},\ \bibinfo {pages} {325} (\bibinfo {year}
  {1999})}\BibitemShut {NoStop}%
\bibitem [{\citenamefont {Wu}\ \emph {et~al.}(2017)\citenamefont {Wu},
  \citenamefont {Cheng}, \citenamefont {Tao}, \citenamefont {Daemen},
  \citenamefont {Foo}, \citenamefont {Nguyen}, \citenamefont {Zhang},
  \citenamefont {Beste},\ and\ \citenamefont {Ramirez-Cuesta}}]{Wu2017JACS}%
  \BibitemOpen
  \bibfield  {author} {\bibinfo {author} {\bibfnamefont {Z.}~\bibnamefont
  {Wu}}, \bibinfo {author} {\bibfnamefont {Y.}~\bibnamefont {Cheng}}, \bibinfo
  {author} {\bibfnamefont {F.}~\bibnamefont {Tao}}, \bibinfo {author}
  {\bibfnamefont {L.}~\bibnamefont {Daemen}}, \bibinfo {author} {\bibfnamefont
  {G.~S.}\ \bibnamefont {Foo}}, \bibinfo {author} {\bibfnamefont
  {L.}~\bibnamefont {Nguyen}}, \bibinfo {author} {\bibfnamefont
  {X.}~\bibnamefont {Zhang}}, \bibinfo {author} {\bibfnamefont
  {A.}~\bibnamefont {Beste}},\ and\ \bibinfo {author} {\bibfnamefont {A.~J.}\
  \bibnamefont {Ramirez-Cuesta}},\ }\bibfield  {title} {\bibinfo {title}
  {{Direct Neutron Spectroscopy Observation of Cerium Hydride Species on a
  Cerium Oxide Catalyst}},\ }\href {https://doi.org/10.1021/jacs.7b05492}
  {\bibfield  {journal} {\bibinfo  {journal} {J. Am. Chem. Soc.}\ }\textbf
  {\bibinfo {volume} {139}},\ \bibinfo {pages} {9721} (\bibinfo {year}
  {2017})}\BibitemShut {NoStop}%
\bibitem [{\citenamefont {Shannon}(1976)}]{Shannon1976}%
  \BibitemOpen
  \bibfield  {author} {\bibinfo {author} {\bibfnamefont {R.~D.}\ \bibnamefont
  {Shannon}},\ }\bibfield  {title} {\bibinfo {title} {{Revised effective ionic
  radii and systematic studies of interatomic distances in halides and
  chalcogenides}},\ }\href {https://doi.org/10.1107/S0567739476001551}
  {\bibfield  {journal} {\bibinfo  {journal} {Acta Crystallogr., Sect. A:
  Found. Crystallogr.}\ }\textbf {\bibinfo {volume} {32}},\ \bibinfo {pages}
  {751} (\bibinfo {year} {1976})}\BibitemShut {NoStop}%
\end{thebibliography}
%

\end{document}